\documentclass[a4paper]{jpconf}
\usepackage{graphicx}    
\usepackage{amsmath}
\usepackage{esint}
\usepackage{array}       
\usepackage{verbatim}    
\usepackage{color}
\usepackage{colortbl}
\usepackage{listings}
\usepackage{epstopdf}
\usepackage{subfigure}
\usepackage{rotating}
\usepackage[percent]{overpic}
\usepackage{wrapfig}
\usepackage{url}
\newcommand{\boxedblue}[1]{\textcolor{blue}{\fbox{\normalcolor $\displaystyle #1$}}}
\newcommand{\boxedred}[1]{\textcolor{red}{\fbox{\normalcolor $\displaystyle #1$}}}
\newcommand{\boxedgreen}[1]{\textcolor{green}{\fbox{\normalcolor $\displaystyle #1$}}}

\newtheorem{thm}{Theorem}
\newtheorem{deff}{Definition}

\definecolor{darkred}{RGB}{175,0,0}

\begin{document}
\title{Advancing parabolic operators in thermodynamic MHD models:  Explicit super time-stepping versus implicit schemes with Krylov solvers}
\author{R. M. Caplan, Z. Miki{\'c}, J. A. Linker, and R. Lionello}
\address{Predictive Science Inc., 9990 Mesa Rim Road Suite 170, San Diego, CA  92121}
\ead{caplanr@predsci.com, mikicz@predsci.com, linkerj@predsci.com, lionel@predsci.com}

\begin{abstract}
We explore the performance and advantages/disadvantages of using unconditionally stable explicit super time-stepping (STS) algorithms versus implicit schemes with Krylov solvers for integrating parabolic operators in thermodynamic 
MHD models of the solar corona.  
Specifically, we compare the second-order Runge-Kutta Legendre (RKL2) STS method with the implicit backward Euler scheme computed using the preconditioned conjugate gradient (PCG) solver with both a point-Jacobi and a non-overlapping domain decomposition ILU0 preconditioner.
The algorithms are used to integrate anisotropic Spitzer thermal conduction and artificial kinematic viscosity at time-steps much larger than classic explicit stability criteria allow. 
A key component of the comparison is the use of an established MHD model (MAS) to compute a real-world simulation on a large HPC cluster.  Special attention is placed on the parallel scaling of the algorithms.  
It is shown that, for a specific problem and model, the RKL2 method is comparable or surpasses the implicit method with PCG solvers in performance and scaling, but suffers from some accuracy limitations.  These limitations, and the applicability of RKL methods are briefly discussed.
\end{abstract}

\section{Introduction}
\label{sec:intro}
Complex physical systems often have processes that act on widely different time-scales. For magnetohydrodynamic (MHD) models of the Solar corona, these processes include plasma flow, slow and fast magneto-sonic waves, resistivity, kinematic viscosity, and thermal conduction \cite{2014_Priest_BOOK}.   When implementing the models with numerical methods, accurately capturing the dynamics of these wide-ranging time-scales can make the simulations computationally intractable due to the fine time-resolutions required.  Also, when using classic explicit finite difference algorithms (such as the Euler method), there exists numerical time-step stability restrictions (typically a function of grid cell size) that can limit the time-step further.  

In order to make the simulations computationally feasible, they need to be integrated at time-steps which exceed the fastest time-scales.  This is possible because often, a system can be integrated without accurately capturing the detailed evolution of the faster time-scales and still obtain useful solutions.  However, integrating at large time-steps can violate the explicit numerical stability requirements, and if too large, can cause the model to loose convergence and become very inaccurate.  Therefore, accuracy considerations must be taken into account when integrating past the fast time-scales, and numerical methods that can overcome the explicit limits are required.

In our model, we adhere to the flow speed time-scale, but exceed the remaining time-scales through a variety of methodologies (see Sec.~\ref{sec:mas}).  The focus of this paper is comparing methods to exceed the explicit numerical time-step restrictions of parabolic operators.  To do this, the operators are first isolated from the rest of the terms in their equations of motion through operator-splitting \cite{OPSPLIT} and are written as
\begin{equation}
\label{eq:pde}
\frac{{ \partial u}}{\partial t}=F({u}),
\end{equation}
where $F$ is the parabolic operator acting on the variable ${u}$.

There exists a number of methodologies which allow for the efficient integration of Eq.~(\ref{eq:pde}) including implicit finite difference methods (e.g. backward-Euler, Crank-Nicholson, etc.), Saul'yev unconditionally-stable explicit schemes \cite{USE_2007,SAUL_CHIN_2014}, and super time-stepping schemes (also known as extended-stability Runge-Kutta schemes) \cite{verwer1996explicit}.  In some cases, smaller time-scale non-parabolic operators yielding the effect of the original operator can be found (e.g. see Ref~\cite{2012_IMPmhd_TCwave} where a wave formulation of thermal conduction is used).  Implicit schemes typically lead to large systems of equations at each time-step that are often solved using Krylov subspace methods (in the linear case), or Newton-Krylov methods (for nonlinear operators) \cite{NSOLI}.

This paper seeks to compare the basic advantages, disadvantages, performance, and parallel scaling of two classes of methods (implicit and super time-stepping) by using a representative algorithm of each class.  The chosen algorithms are 1) The implicit Backwards-Euler scheme solved with the Preconditioned Conjugate Gradient method (BE+PCG) and 2) The explicit second-order Runge-Kutta Legendre (RKL2) super time-stepping method \cite{RKL2_2012, RKL2_2014}.  The criteria for the comparisons consists of looking at accuracy validation, the ease/difficulty of the method's formulation and implementation, the features and limitations of each method, their overall performance (e.g. the number of iterations/sub-steps needed for integrating each large time-step, the computational cost of each iteration/sub-step), and the methods' parallel performance including communication/synchronization requirements and their ability to scale to large numbers of processors.  

The two methods are used to integrate the most time-consuming parabolic operators of our model: Spitzer anisotropic thermal conduction and artificial kinematic viscosity.  Since the PCG method is only applicable to linear operators, the thermal conduction operator is linearized using lagged diffusivity \cite{LAG_DIF} (i.e. we fix the temperature-dependent part of the diffusion coefficient using the previous time-step's temperature values).

A previous comparison between PCG and the factored RKC STS method \cite{RKC_1996} for a 1D diffusion test case using the RAMSES code was done in Ref.~\cite{PCGDvsRKC}, where the authors found the STS method to be much faster (and more accurate) than their PCG method.  Comparisons between RKL2 and a Newton-Krylov method using GMRES and with a multigrid solver was done for several test problems by the authors of the RKL method in Ref.~\cite{RKL2_2014}, where similar speed improvement results were found.  A key element to the current comparison is the use of a large, full production simulation to test the methods.

Since it is not feasible to cover every possible combination of problem size, problem type, HPC environment, etc., our results presented here are not to be taken as generally applicable in all situations, but rather as a single real-world example to help gain some flavor of the advantages/limitations of using these two very different methodologies.  The production run used here is a full-scale coronal simulation (described in Sec.~\ref{sec:testcase}) computed using two state-of-the-art HPC systems (San Diego Supercomputing Center's Comet and Texas Advanced Computing Center's Stampede machines).  

The paper is organized as follows:  The two methods are described in Sec.~\ref{sec:BEPCG} (BE+PCG) and Sec.~\ref{sec:rkl2} (RKL2).  Sec.~\ref{sec:mas} describes the MAS thermodynamic MHD model used to simulate the production run described in Sec.~\ref{sec:testcase}. The validation of the two methods is discussed in Sec.~\ref{sec:valid} including important applicability concerns for RKL methods.  The timing results and analysis are described in Sec.~\ref{sec:timings}, and we conclude in Sec.~\ref{sec:conclusion}.

\section{Implicit Backward Euler Solved with Preconditioned Conjugate Gradient}
\label{sec:BEPCG}
There exists a wide range of implicit discretization schemes to integrate Eq.~(\ref{eq:pde}) and numerous methods to solve the resulting system of equations.  For the present context, we limit ourselves to the backward Euler (BE) scheme solved with a preconditioned conjugate gradient (PCG) sparse solver (we denote the full method as BE+PCG).  The BE+PCG method is chosen as it is commonly used and is the current method applied in the MAS code.  Although the method is well-known, we review it in this section in order to highlight its advantages and disadvantages, especially in relation to parallel performance. 
\subsection{Implicit Backward Euler}
One of the easiest-to-implement implicit methods for integrating Eq.~(\ref{eq:pde}) is the BE scheme given by
\begin{equation}
\label{eq:BE}
\frac{{ u^{n+1}}-{ u^n}}{\Delta t}=F({ u^{n+1}}),
\end{equation}
where $\Delta t=t^{n+1}-t^n$ is the chosen time-step and $u^n=u(t^n,{\bf x})$.  If one assumes that $F$ is a linear self-adjoint operator, then Eq.~(\ref{eq:BE}) can be written as
\begin{equation}
\label{eq:BE_lin}
\frac{{ u^{n+1}}-{ u^n}}{\Delta t}={\bf M}\,u^{n+1},
\end{equation}
where ${\bf M}$ is a symmetric positive-definite matrix.  Given the solution at step $n$ ($u^n$), Eq.~(\ref{eq:BE_lin}) leads to a system of equations,
\begin{equation}
\label{eq:BE_linsolve}
{\bf A}\,u^{n+1}=u^n,
\end{equation}
where ${\bf A}=1-\Delta t\,{\bf M}$, which, when solved yields $u^{n+1}$.  The BE method is a low-order method ($O(\Delta t)$ accurate in time), but because it is unconditionally stable, L-stable (i.e. its amplification factor tends to 0 as the wave mode number tends to $\infty$), and very simple to implement, it remains a useful and wide-spread method.

\subsection{Preconditioned Conjugate Gradient}
The matrix ${\bf A}$ of Eq.~(\ref{eq:BE_linsolve}) is often large and very sparse.  A common method for solving such sparse systems are Krylov subspace iterative solvers (see Ref.~\cite{IterativeMethods_SAAD_Book} for a detailed review), and in the case of symmetric operators, the Conjugate Gradient (CG) method.  In order to allow the CG method to converge to the solution efficiently, the system is preconditioned with an approximate inverse of the matrix ${\bf P}^{-1} \approx {\bf A}^{-1}$, leading to the preconditioned-CG method (PCG).  The application of the preconditioner (${\bf P}^{-1}\,x$) is almost always performed without explicitly formulating the inverse matrix ${\bf P}^{-1}$. The PCG algorithm as used in MAS is shown in Fig.~\ref{fig:pcg} where we have highlighted the expressions that require inter-process communication with a green box to denote local communication/synchronization (stencil interchange) and a red box to denote global communication/synchronization (global summation for the dot products).  While the basics of the PCG method are relatively simple to implement, it is difficult to properly choose optimal convergence criteria/values \cite{PCGstopcrit} and, even more so, an efficient preconditioner ${\bf P}$.
\newpage
\subsection{Preconditioners}
\label{sec:PC}
\begin{wrapfigure}{r}{0.5\textwidth}
\[
\begin{array}{cc}
\begin{array}{l}
x_0=u^n\notag \\
r_0=b-{\bf A}\,x_0 \notag \\
z_0={\bf P}^{-1}\,r_0 \notag \\
p_0=z_0 \notag \\
r_r=r_0\cdot z_0 \notag  
\end{array}
&
\begin{array}{l}
{\tt do\; k=0:k_{\mbox{\small max}}}\notag  \\
  \;\;\; y_k=\boxedgreen{{\bf A}\,p_k}\notag  \\
  \;\;\; \alpha_k=r_r/\boxedred{(p_k\cdot y_k)}\notag \\
  \;\;\; x_{k+1}=x_k+\alpha_k\,p_k\notag \\
  \;\;\; r_{k+1}=r_k-\alpha_k\,y_k\notag \\
  \;\;\; z_{k+1}={\bf P}^{-1}\,r_{k+1}\notag \\
  \;\;\; r_{\mbox{\small old}}=r_r\notag \\
  \;\;\; r_r=\boxedred{r_{k+1}\cdot z_{k+1}}\notag \\
  \;\;\; \boxedred{{\tt Check}\,r_r\,{\tt for\, convergence}}\notag \\
  \;\;\; \beta_k=r_r/r_{\mbox{\small old}}  \notag \\
  \;\;\; p_{k+1}=\beta_k\,p_k + z_k,  \notag \\
{\tt enddo \qquad \qquad}\notag \\
u^{n+1}=x_{k+1}
\end{array}
\end{array}
\]
\caption{The PCG algorithm with inter-processor communication terms boxed (red: global, green: point-to-point)\label{fig:pcg}}
\end{wrapfigure}
A major difficulty with using the PCG method is selecting and computing an inexpensive yet efficient preconditioner (PC).   Preconditioning is a very complex topic with many decades of research and a multitude of methods \cite{PC_iterative_survey}.  The use of any method has many weighing factors such as the amount of computational work, complexity of implementation, ability to parallelize efficiently, applicability to the linear solver and matrix representation, as well as problem-specific considerations that can reduce the usefulness of the PC (and in some cases can even cause the PCs to break down \cite{ILU_breakdown}).  Another difficulty is that certain PCs which work very well to improve convergence, can be very difficult to implement efficiently (or at all) on large HPC systems. 
 
For the comparisons in this paper, we use two relatively simple communication-free PCs: 1) A Point-Jacobi/diagonal scaling (PC1), which simply uses the inverse of the diagonal of ${\bf A}$.  The PC1 is very inexpensive to formulate and apply, but is limited in its effectiveness.
2) A non-overlapping domain decomposition with zero-fill incomplete LU factorization (PC2), which is more expensive to formulate and apply than PC1, but is also much more effective at reducing iterations.  One drawback of PC2 is that, due to the non-overlapping implementation (used to avoid adverse scaling performance), it becomes less effective as the number of processors increases.  In fact, in the non-practical upper limit of one processor core per grid point, PC2 becomes equivalent to PC1.  However, for realistic uses, the degradation in convergence remains small for reasonable grid points per processor ratios (see Sec.~\ref{sec:timings}).  Pseudo-code descriptions of the implementation for the initial formulation and subsequent application of both PCs is described in Appendix~A.

Many new methods/preconditioners such as Geometric/Algebraic multigrid \cite{2000_Briggs_Multigrid} and inter-processor overlapping PCs \cite{shen2002parallel} exist, which would yield different performance and scaling results than presented here.  However, such methods are often substantially more difficult to implement and parallelize (especially into legacy code) and often have trade-offs in effectiveness and parallel scaling that are hard to predict.  
 
\subsection{Advantages and disadvantages of the BE+PCG method}
Some advantages of using the BE+PCG approach are that; (a) it is widely used, robust, and has well-known convergence properties; (b) the BE scheme is L-stable, which efficiently damps out high wave modes (see Sec.~\ref{sec:valid}); (c) it can be used for slightly non-symmetric operators \cite{PCG4nonsym}; (d) its basic form is fairly straightforward to implement; (d) there exists many available `black-box' implementations, and (e) it can be efficient given a good preconditioner and implementation. 

However, the BE+PCG approach does suffer from some important disadvantages.  One limitation in using the PCG method (or any Krylov solver) is that it requires a linear operator.  Often, (as in the case of thermal conduction) the parabolic terms have non-linear coefficients and/or are desired to be flux-limited.  In order to do this, using BE, one would have to employ a nonlinear system solver such as the popular Newton-Krylov methods \cite{NSOLI} which can be computationally expensive.  Efficiently porting the PCG method to accelerator hardware is a formidable challenge and, as mentioned in the previous section, choosing an optimal stopping criteria and efficient preconditoner is often very difficult.  

Running PCG solvers on large HPC systems exposes other disadvantages.  The global communication/synchronization requirement of the dot-products in the PCG algorithm can adversely effect the scaling of the method, especially in the presence of (even small) load imbalance (see Sec.~\ref{sec:timings}).  There are ongoing efforts to mitigate this problem \cite{PCG_NONBLOCK}, but they have yet to show substantial improvement.  This is problematic for domain decomposed problems (such as ours) because the grid resolution in each dimension does not typically divide evenly into the number of processors assigned to that dimension, leading to some sub-domains having more grid points than others.  Therefore, as the problem is spread to more and more processors, the load imbalance will increase (see Fig.~\ref{fig:iters} in Sec.~\ref{sec:timings}).

\section{Explicit Runge-Kutta Legendre Super Time-stepping scheme}
\label{sec:rkl2}
A relatively recent class of schemes for integrating parabolic operators explicitly without numerical restriction on the time-step size are `super time-stepping'/`extended stability' (STS) methods.  The main idea behind them is to use a Runge-Kutta multi-stage scheme with stages added for stability rather than accuracy.  For a given desired time-step, a number of stages/sub-steps ($s$) and specialized coefficients for each stage are found such that the overall advance is stable.  
\begin{wrapfigure}{l}{0.5\textwidth}
\[
\begin{array}{l}
M_0=\boxedgreen{{\bf M}\,u^n} \\
u_1 = u^n + \tilde{\mu_1}\,\Delta t\,M_u  \\
{\tt do\; k=2:s} \\
\;\;\; u_k = \mu_j\,u_{k-1} + \nu_j\,u_{k-2}+(1-\mu_k-\nu_k)\,u^n  \\
\qquad \;\;\;  + \tilde{\mu_k}\,\Delta t\,\boxedgreen{{\bf M}\,u_{k-1}}+(b_k-1)\,\tilde{\mu_k}\,\Delta t\,M_0 \\
{\tt enddo} \\
u^{n+1} = u_s, 
\end{array} \notag
\]
\caption{The RKL2 algorithm with inter-processor communication terms boxed (green: point-to-point)\label{fig:rkl2}}
\end{wrapfigure}
STS methods were made popular as a practical numerical method in 1996 \cite{verwer1996explicit} but remained relatively unknown and unused until recent years where interest in the methods and their development have rapidly increased.  STS methods are based on either Chebychev polynomials (RKC) \cite{RKC_1996} or Legendre polynomials (RKL) \cite{RKL2_2012,RKL2_2014}, and are either formulated as factored \cite{FRKC_highorder} or recursive \cite{verwer1996explicit} methods.

The RKC STS method has been used successfully for parabolic operators in MHD codes \cite{RKC1_2007,FLASH_STS,PLUTO1} and the RKL method is currently being implemented into the Lare3D MHD code \cite{LARE3D_2001,LARE3D_code} and has been shown to be accurate and efficient for some coronal loop simulations \cite{LOOPS_RKL}.   Ref.~\cite{RKL2_2014} has shown that for nonuniform/nonlinear diffusion coefficients (such as ours), the RKL STS method has more desirable properties than the RKC. We therefore use the RKL method, and for accuracy and robustness, we use the second-order accurate version (RKL2) \cite{RKL2_2014}. 
 
The algorithm for the RKL2 scheme is shown in Fig.~\ref{fig:rkl2}.  The pre-computed coefficients for each stage are defined as
\begin{alignat}{3}
b_0&=b_1=b_2=\frac{1}{3},\qquad                    \;\;\;\;\;\; b_k = \frac{k^2+k-2}{2\,k\,(k+1)}, \notag \\
\tilde{\mu_1} &= \frac{4}{3\,(s^2+s-2)}, \qquad   \;\;\; \mu_k = \frac{2\,k-1}{k}\,\frac{b_k}{b_{k-1}}, \notag  \\
\tilde{\mu_k} &=\frac{4\,(2\,k-1)}{k\,(s^2+s-2)}\,\frac{b_k}{b_{k-1}}, \;\;\; \nu_k = -\frac{k-1}{k}\,\frac{b_k}{b_{k-2}}.\notag 
\end{alignat}
The number of stages/sub-steps needed for stability (given a desired $\Delta t$) is given by
\begin{equation}
\label{eq:rkl2_s}
s=\frac{1}{2}\,\left\lceil\sqrt{9+16\,\frac{\Delta t}{\Delta t_{\mbox{\tiny Euler}}}}-1\right\rceil,
\end{equation}
where $\Delta t_{\mbox{\tiny Euler}}$ is the explicit Euler time-step stability limit (for a simple and fast way to calculate $\Delta t_{\mbox{\tiny Euler}}$ that avoids too much underestimating, see Appendix~B).  For better accuracy, the authors of Ref.~\cite{RKL2_2014} recommend always ensuring $s$ is odd (although in our experience, when using more sub-steps than Eq.~(\ref{eq:rkl2_s}), this is unnecessary).

\begin{wrapfigure}{l}{0.5\textwidth}
  \centering
  \includegraphics[width=0.48\textwidth]{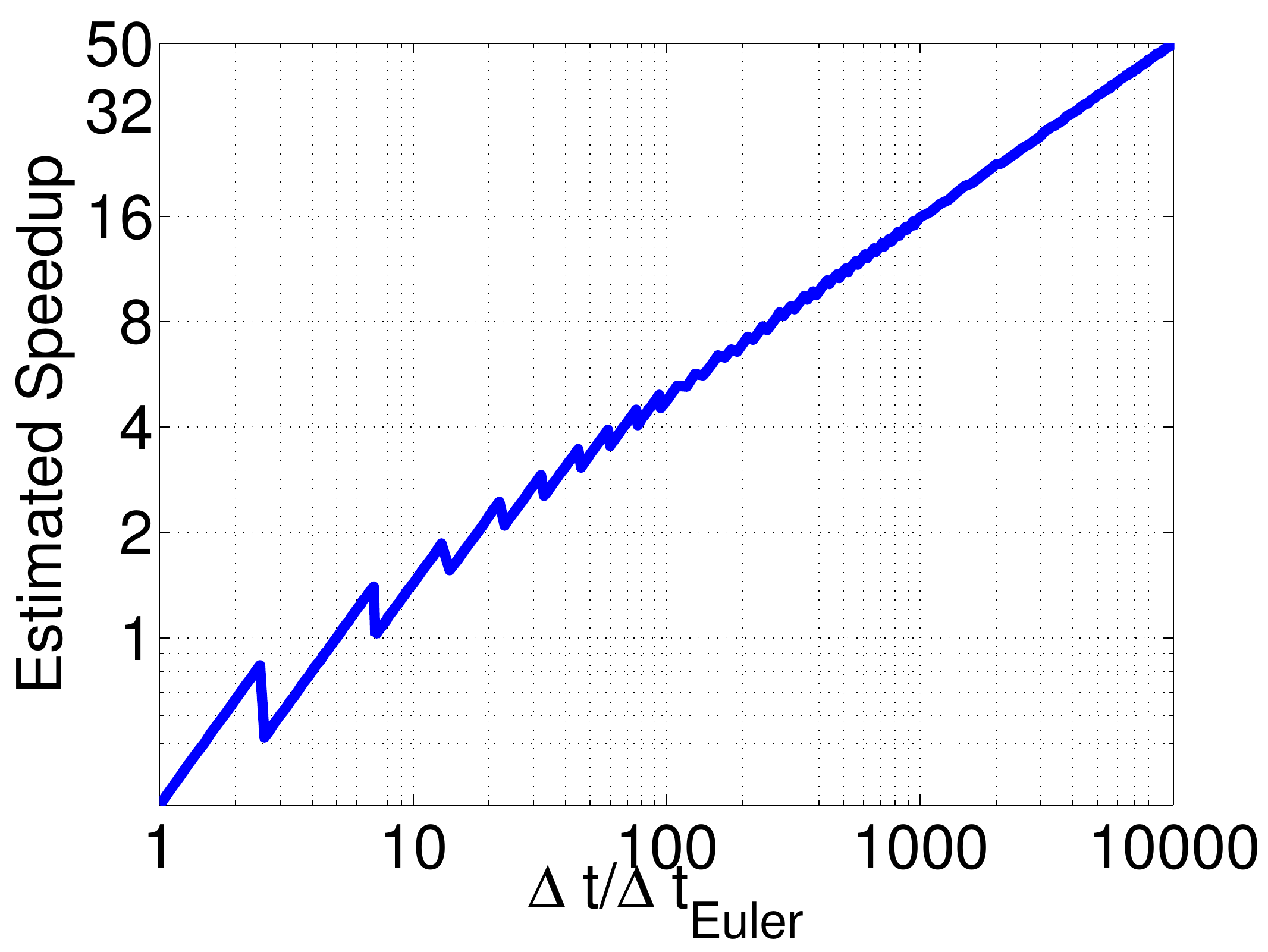}
  \caption{Estimated speedup of using the RKL2 versus the explicit Euler method assuming one Euler step has the same computational cost as one RKL2 sub-step.  The jaggedness of the plot is due to the discrete nature of Eq.~(\ref{eq:rkl2_s}). \label{fig:rkl2speedup}}
\end{wrapfigure}  
Since $\Delta t_{\mbox{\tiny Euler}}\propto \Delta x^2$, the number of sub-steps follows $s\propto 1/\Delta x$ whereas sub-cycling the Euler method would take $\propto 1/\Delta x^2$ steps.  Therefore, the RKL2 scheme requires many less steps than the Euler method and the larger the overall time-step is, the more speedup the RKL2 method exhibits (see Fig.~\ref{fig:rkl2speedup}).  As in describing the PCG solver, in Fig.~(\ref{fig:rkl2}) we have highlighted the terms in the RKL2 method that require parallel communication.   Unlike the PCG solver, the RKL2 scheme only has local communication/synchronization, so is expected to scale much better in the presence of load imbalance and system performance fluctuations (see Sec.~\ref{sec:timings} where this is indeed the case).

\subsection{Advantages and disadvantages of RKL2}
Some of the main advantages of the RKL2 method (and other STS methods) is that it is very easy to implement (both in a matrix and matrix-free manner), lends itself to easy implementation of directive-based parallelism (such as OpenMP/OpenACC), does not require global synchronizations/communications, and can be used with nonlinear operators/flux-limiting.  Its performance gains can vary, but in cases where the accuracy of the fast-time scale of the operator is not needed, it can have substantial speedup over standard explicit methods, and, as we will show in Sec.~\ref{sec:timings}, can sometimes outperform implicit methods.

One disadvantage of the RKL2 method is that it is very new with very few real-world implementations to test its robustness in a variety of settings.  Also, in some cases, the number of sub-steps/stages can become very large, in which case implicit methods may outperform it.  One additional aspect of the RKL2 scheme is that, as will be discussed in more detail in Sec.~\ref{sec:valid}, its amplification factor does not generally monotonically decrease for increasing wave mode numbers, and is quite high for the highest modes.  This can cause solution artifacts when using only a few total time-steps on solutions containing high wave mode structures, and, when using large time-steps in coupled systems, may fail to damp the high wave modes.

\section{The MAS Thermodynamic MHD Model}
\label{sec:mas}
The MHD model we use to test the RKL2 and BE+PCG methods is the Magnetohydrodynamic Algorithm outside a Sphere (MAS) code \cite{MAS}.  MAS integrates the time-dependent resistive thermodynamic MHD equations in spherical coordinates and has been used extensively in models of coronal structure \cite{mikic99a,linker99a,MAS,downsetal2013}, coronal dynamics \cite{lionello05a,lionello06a,linkeretal2011} and coronal mass ejections \cite{linker03a,lionelloetal2013}.  It is also the primary MHD model in CORHEL (Corona-Heliosphere), a suite of models \cite{rileyetal2012} for describing the solar corona and inner heliosphere that is available for public use at NASA's CCMC \footnote{\url{http://ccmc.gsfc.nasa.gov/}}.  In this section we describe the model equations and numerical methods used in MAS as they are configured for our comparisons (see Sec.~\ref{sec:testcase}).

\subsection{Model equations}
\label{sec:masmodel}
In MAS, the one-fluid ideal MHD equations \cite{2014_Priest_BOOK} are extended to include additional physics with a wide variety of models options and parameters.  For the problem used in this paper, the governing equations take the form:
\begin{alignat}{2}
\frac{\partial {\bf A}}{\partial t} &={\bf v}\times \left(\nabla\times {\bf A}\right)-\boxed{\frac{c^2\,\eta}{4\,\pi}\,\nabla \times \nabla \times {\bf A}} \label{eq:mhd:a},
\\
\frac{\partial \rho}{\partial t} &=-\nabla\cdot(\rho\,{\bf v}),
\\
\frac{\partial T}{\partial t} &=- \nabla\cdot(T{\bf v})
-(\gamma - 2)\left(T\,\nabla\cdot{\bf v}\right)-\frac{1}{2}\,\nabla\cdot\left(f_{\mbox{\tiny nc}}(r)\,T\,{\bf v}\,{\bf \hat b}{\bf \hat b}\right) \label{eq:mhd:e1}
\\
&+\boxedblue{\left.\frac{(\gamma-1)}{2\,k}\frac{m_p}{\rho}\right[\nabla\cdot\left(\beta_{\mbox{\tiny Tcut}}(T)\,f_{\mbox{\tiny c}}(r)\,\kappa_0\,T^{5/2}\,{\bf \hat b}{\bf \hat b}\cdot\nabla T\right)}\left. - \frac{\rho^2}{m_p^2}\,\frac{Q(T)}{\beta_{\mbox{\tiny Tcut}}(T)} + H\right], \notag
\\
\frac{\partial \epsilon_{+}}{\partial t} &=-\nabla\cdot \left( \epsilon_{+}\left[{\bf v}+v_A\,{\bf \hat b}\right]\right) - \frac{1}{2}\,\epsilon_{+}\,\nabla\cdot {\bf v}, \label{eq:mhd:wkb1}
\\
\frac{\partial \epsilon_{-}}{\partial t} &=-\nabla\cdot \left( \epsilon_{-}\left[{\bf v}-v_A\,{\bf \hat b}\right]\right) - \frac{1}{2}\,\epsilon_{-}\,\nabla\cdot {\bf v}, \label{eq:mhd:wkb2}
\\
\frac{\partial{\bf v}}{\partial t}& = -{\bf v}\cdot \nabla\,{\bf v} + \frac{1}{\rho}\left[ \frac{1}{c}{\bf J} \times {\bf B} - \nabla p - \nabla\left(\frac{\epsilon_{+}+\epsilon_{-}}{2}\right) + \rho\,{\bf g}\right]  \label{eq:mhd:v}
\\
&+ \boxedblue{\frac{1}{\rho}\,\nabla\cdot(\nu \rho \nabla {\bf v})}  + \frac{1}{\rho}\,\nabla\cdot\left(S\,\rho\,\nabla{\frac{\partial \bf v}{\partial t}}\right), \notag
\end{alignat}
where ${\bf A}$ is the magnetic vector potential, ${\bf B}=\nabla\times {\bf A}$ is the magnetic field, ${\bf J}=\frac{c}{4\,\pi}\nabla\times {\bf B}$ is the current density, $\rho$ is the plasma density, $T$ is the temperature, $p=2\,k\,T\,\rho/m_p$ is the plasma pressure, {\bf v} is the plasma velocity, ${\bf \hat b}=|{\bf B}|/{\bf B}$ is the normalized direction of the magnetic field, $c$ is the speed of light in a vacuum, $\gamma=5/3$ is the adiabatic index, $m_p$ is the proton mass, $k$ is Boltzman's constant, $\kappa_0$ is the Spitzer's constant, $f_{\mbox{\tiny c}}(r)=0.5\,(1-\mbox{tanh}[(r-10R_{\odot})/0.5\,R_{\odot}])$ is a profile that limits the radial extend that collisional thermal conduction is active, $f_{\mbox{\tiny nc}}(r)=1-f_{\mbox{\tiny c}}(r)$ is the equivalent profile for collisionless thermal conduction (the last term of Eq.~(\ref{eq:mhd:e1}) \cite{CHC_1976}), $Q(T)$ is a radiative loss function (here we select the piece-wise exponential approximation of Ref.~\cite{RADLOSS_ATHAY}), $H$ is a series of combined empirical heating models that are a function of ${\bf B}$ \cite{MAS}, $v_A=\sqrt{|{\bf B}|^2/4\,\pi\,\rho}$ is the Alfv\'en wave speed, ${\bf g}=-g_0\,R_{\odot}^2\,{\bf \hat r}/r^2$ is the gravitational force, and the last expression in Eq.~(\ref{eq:mhd:v}) is a semi-implicit term described in the next section.  Eqs.~(\ref{eq:mhd:wkb1}) and (\ref{eq:mhd:wkb2}) are a WKB approximation for Alfven wave pressure advance \cite{mikic99a} where $\epsilon_{+}$ and $\epsilon_{-}$ are the forward and backward Alfv\'en wave pressures.

The resistivity term in Eq.~(\ref{eq:mhd:a}) is added in order to dissipate structures that cannot be resolved since they are smaller than the cell size.  For the current simulations, we use a constant value of $\eta=4.678\times 10^{-8}\,\mbox{s}$ corresponding to a magnetic diffusion coefficient of $\sim 10^{12}\,\mbox{cm}^2/\mbox{s}$.  While this is much larger than the typical chromospheric ($\sim 10^7\,\mbox{cm}^2/\mbox{s}$) and coronal ($\sim 10^4 \,\mbox{cm}^2/\mbox{s}$) values \cite{2014_Priest_BOOK}, the corresponding magnetic diffusion time scale remains much smaller than the typical Alfv\'en travel time (yielding Lundquist numbers of $\sim 10^6$).  Therefore, although sub-grid level structure is diffused, the overall solution is not affected substantially.  The artificial viscosity term of Eq.~(\ref{eq:mhd:v}) is added for a similar reason, in that it serves to dissipate unresolvable stricture.  For our problem, we use a kinematic viscosity value of $\nu=6.7\times 10^{15}\,\mbox{cm}^2/\mbox{s}$, which, while very large, yields a diffusion time-scale for velocity that is still $500$ times smaller than the typical Alfv\'en travel time.

The function $\beta_{\mbox{\tiny Tcut}}(T)$ is a cut-off function that sets a minimum temperature that can be used for the thermal conduction coefficient as $\beta_{\mbox{\tiny Tcut}}=(T/T_{\mbox{\tiny cut}})^{5/2}$ for $T<T_{\mbox{\tiny cut}}$ and $\beta_{\mbox{\tiny Tcut}}=1$ for $T \ge T_{\mbox{\tiny cut}}$, where $T_{\mbox{\tiny cut}}=5\times 10^5 K$.  Applying this cut-off function serves to broaden the width of the transition region (to make it more numerically resolvable) in such a way that it has a minimal effect on the global coronal solution \cite{TCmod2001}.

In describing the MHD equations above, we have boxed the parabolic terms, where those used to test the two methods are colored blue (the resistivity term uses a very small number of PCG iterations to solve, and is therefore excluded from the method comparisons). 

\subsection{Numerical methods}
\label{sec:masnum}
The MAS code is written in FORTRAN90 and parallelized using MPI.  It computes the MHD equations using finite difference on a logically rectangular non-uniform spherical grid.  The non-uniformity of the grid allows MAS to efficiently resolve small-scale structures such as the transition region and active regions, while allowing for coarser grid points over the global scale.  Each component of the field vectors are staggered which, combined with the use of ${\bf A}$ as a primitive, ensures that $\nabla\cdot{\bf B}=0$ exactly.  Advective terms are differenced using first-order upwinding, while parabolic and gradient terms are centrally differenced.  The non-uniformity of the grid is implemented into the discretizations using the standard method `A' of Ref.~\cite{NUG_PLAY_1992}.  For the anisotropic thermal conduction operator, we use the `standard' scheme of Ref.~\cite{TCO_STD}.  The viscosity operator is written in a self-adjoint form (excluding boundary conditions) that leads to a nearly-symmetric solver matrix.  

The code uses an adaptive time-step that conforms to the plasma flow CFL condition.  As this is often a much slower time-scale than the magnetosonic waves, a semi-implicit term is added to Eq.~(\ref{eq:mhd:v}) that stabilizes the algorithm for time-steps larger than the fast magneto-sonic wave limit.  The semi-implicit factor is given as $S=(\Delta t^2\,\tilde k^2)^{-1}\,(C_w^2/(1-C_f)^2-1)$, where $\tilde k^2$ is the upper limit of the combined inverse grid spacing, $C_f$ is the flow CFL, and $C_w$ is the fast magnetosonic wave CFL \cite{ROBSTB}. 

The temporal differencing uses a predictor-corrector scheme for the advective and semi-implicit operator, where any reactive terms are included in the corrector step.  The parabolic terms are first-order operator split.  The WKB approximate Alfv\'en wave pressure advance of Eqs.~(\ref{eq:mhd:wkb1}) and (\ref{eq:mhd:wkb2}) are sub-cycled at their explicit advective CFL condition based on $v_A$.  

All parabolic operators and the semi-implicit solves are currently computed using the BE+PCG method with PC1 for the resistivity, and PC2 for the remaining solves.  The operator matrices are stored in a DIA sparse storage format \cite{DIACSR} for internal grid points, while the boundary conditions are applied matrix-free.  The PC2 preconditioner LU matrix is stored in the CSR sparse storage format optimized for memory access described in Ref.~\cite{CSRopt}.  In our implementation of the STS method for this comparison, we use the same matrix format used for the solver operators.

\section{Real-world Test Case}
\label{sec:testcase}
\begin{figure}[htbp]
\centering
\includegraphics[height=1.9in]{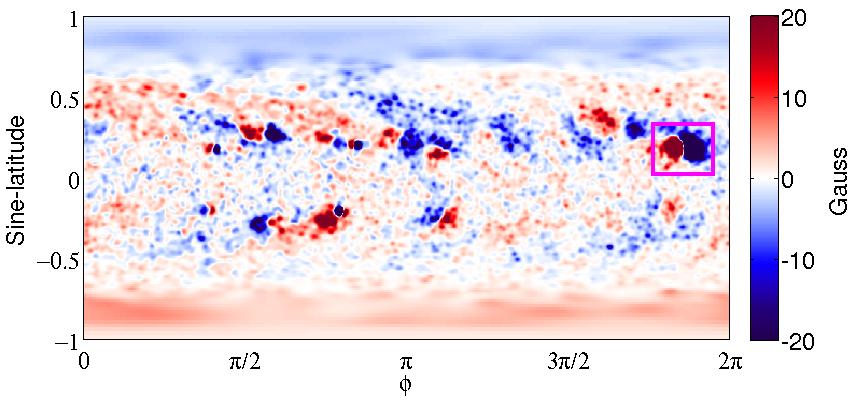}
\includegraphics[height=1.9in]{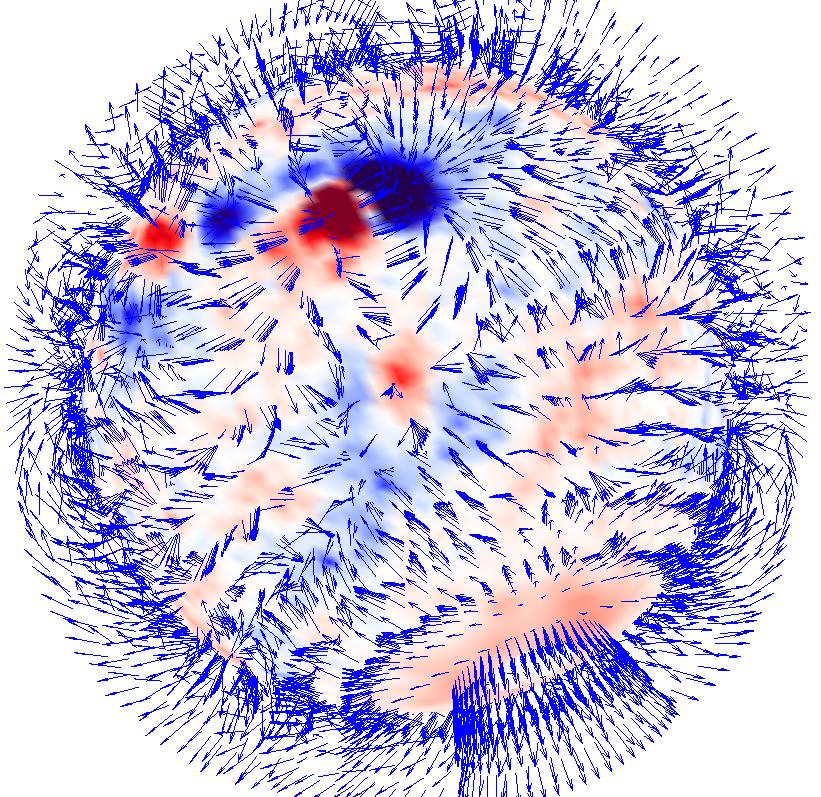}
\caption{Left: Smoothed initial ADAPT map from 07/07/2010 20:00 with added far-side active region (boxed). Right: Sampled unit vectors of initial condition of magnetic field after solving for potential fields (up to $r=1.41\,R_{\odot}$).\label{fig:testcase1}} 
\end{figure}

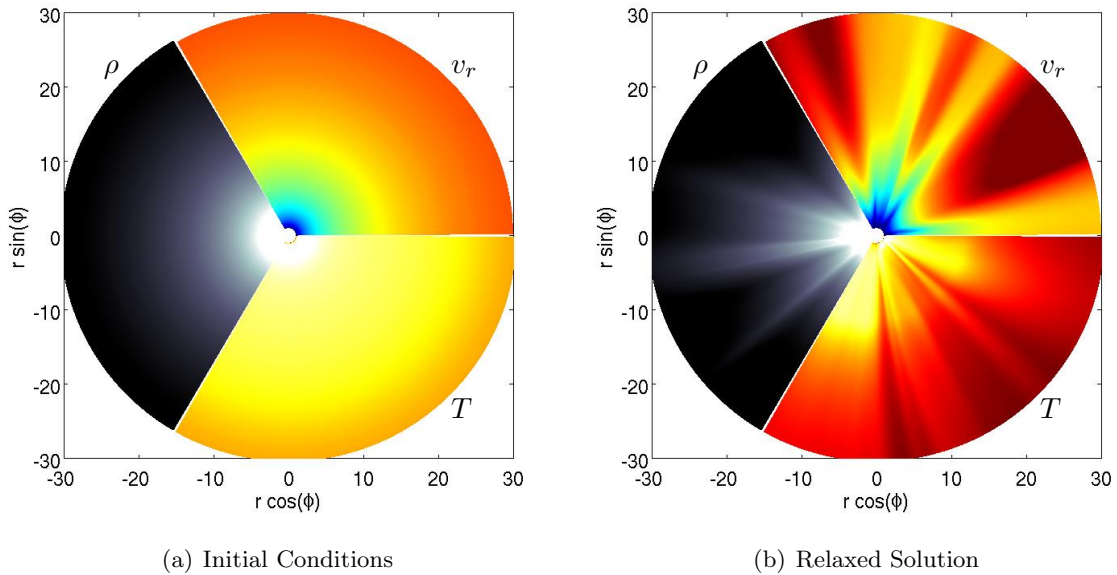
\begin{figure}
\centering
\subfigure[Initial Conditions]{\begin{overpic}[width=3in]{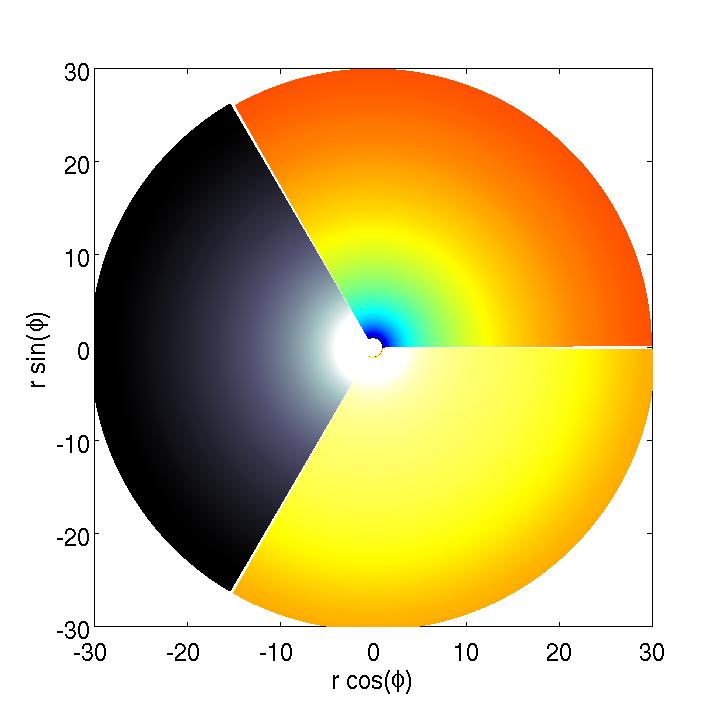}
     \put(20,80){$\rho$}
     \put(80,80){$v_r$}
     \put(80,20){$T$}
\end{overpic}}
\subfigure[Relaxed Solution]{\begin{overpic}[width=3in]{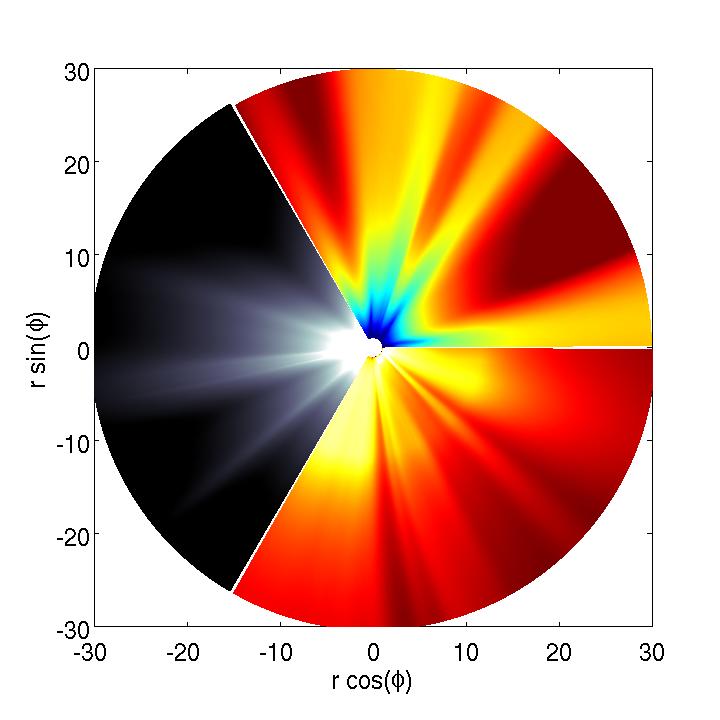}
     \put(20,80){$\rho$}
     \put(80,80){$v_r$}
     \put(80,20){$T$}
\end{overpic}}
\caption{Initial condition (a) and solution at 48 simulation-hours (b) for the coronal relaxation.  Portions within a $r${--}$\phi$ cut at $\theta=\pi$ are shown for $\rho$, $v_r$, and $T$. \label{fig:testcase2}}
\end{figure}

To provide a useful comparison of the methods, we use a production run of the MAS code.  The procedure for producing an MHD model of the solar corona has been described previously \cite{linker99a,MAS}.  In brief, a full-sun map of the radial magnetic field derived from observations is used to specify the boundary conditions at the solar surface.  A potential solution is computed using this boundary condition to supply the initial magnetic field condition (see Fig.~\ref{fig:testcase1}).  A spherically symmetric solar wind solution (see Fig.~\ref{fig:testcase2}) is used as the initial condition for the other variables ($\rho$, $T$, $\vec v$).  This initially non-equilibrium state is advanced in time until a steady-state is approached.  The resulting solutions can be compared against a range of solar and heliospheric observations \cite{rileyetal2012}, including EUV and X-ray emission \cite{MAS}, and white-light eclipse images \cite{ruvsin2010comparing}.  
  
For the case shown here, the surface radial magnetic field is taken from a flux evolution model (ADAPT) that assimilates observational Earth-based magnetograms to create a full-sun magnetogram at a one particular time \cite{ADAPT,ADAPT2}.  Sometimes active regions (AR) can emerge on the far-side of the Sun that are not assimilated into the ADAPT map until they rotate into Earth's view.  In the case being simulated here, such an AR was spotted by the STEREO spacecraft in EUV emission images.  In order to better approximate the synchronic state of the surface field, the far-side AR (as observed later in time) is added into the ADAPT map. The map is then interpolated onto the computational grid and smoothed.  In Fig.~\ref{fig:testcase1}, we show the final form of the ADAPT map used, and the initial condition of ${\bf B}$ derived from the resulting potential field solution.

The full simulation integrates the initial conditions for 48 simulation-hours.  For validation runs (Sec.~\ref{sec:valid}), we integrate from the initial conditions to $t=8$ simulation-hours as this contains the most dynamical/difficult portion of the total run.  For the timing runs, we start with the solution at $t=8$ simulation-hours (computed using MAS's default algorithms) and then integrate using the two methods for another $t=6$ simulation-minutes.

The domain is global and covers a radial range from $1\,R_{\odot}$ to $30\,R_{\odot}$.  It is discretized into a logically-rectangular nonuniform spherical grid ($r,\theta,\phi$).  The grid (summarized in Table~\ref{tab:grid}) is nearly uniform in $\phi$, coarsens slightly towards the pole in $\theta$, and is highly non-uniform in $r$ (in order to better resolve the transition from the upper chromosphere to the corona).
\begin{table}
\centering
$\begin{array}{|l|c|r|r|r|}
\cline{2-5}
\multicolumn{1}{c|}{\;} & \mbox{N} & \mbox{min}(\Delta_i) & \mbox{max}(\Delta_i) & \mbox{max}|1-\Delta_{i+1}/\Delta_i| \\
\hline
r & 181 & ~340 \mbox{km} & ~500,000 \mbox{km} & ~6\% \\
\hline
\theta & 251 & ~0.55^{\circ} & ~1.76^{\circ} & ~3\% \\
\hline
\phi & 602 & ~0.6^{\circ} & ~0.6^{\circ} & 0\% \\
\hline
\end{array}$
\caption{Grid information for the coronal relaxation run.\label{tab:grid}}
\end{table}
The time-step changes over the runs to conform to the flow CFL condition and during the 8-simulation-hour validation simulation, ranged from $\sim2.5$ seconds to $\sim3.4$ seconds with an average step-size of $\sim2.7$ seconds.

In Fig.~\ref{fig:testcase2}, we show some plasma values within an $r$--$\phi$ slice at $\theta=\pi/2$ for the initial condition and the final relaxed corona at $t=48$ simulation-hours.

\subsection{HPC environment}
\label{sec:testcasehpc}
To test the parallel scaling and performance of the methods, we run the simulations on state-of-the-art HPC systems.  In order not to limit results to one system, we use both the Comet machine at the San Diego Supercomputer Center (SDSC) and the Stampede machine at the Texas Advanced Computing Center (TACC).  The hardware and software configuration used for each system are shown in Table~\ref{tab:sysinfo}.
\begin{table}
\centering
\scalebox{0.9}{
$\begin{array}{|l|r|r|}
\multicolumn{1}{c}{\;} & \multicolumn{1}{r}{\bf SDSC\,Comet} & \multicolumn{1}{r}{\bf TACC\,Stampede} \\
\hline
\mbox{Processor Model} & \mbox{Intel Xeon } & \mbox{Intel Xeon }\\
\, & \mbox{E5-2660v3} & \mbox{E5-2680v2} \\ \hline
\mbox{Clock Speed} & 2.5\,\mbox{GHz} & 2.7\,\mbox{GHz}\\ \hline
\mbox{CPU Cores/node} & 24 & 16\\  \hline
\mbox{DRAM/node} & 128\,\mbox{GB} & 32\,\mbox{GB} \\ 
\mbox{(ECC)} & \mbox{DDR4-2133MHz} & \,\mbox{DDR3-1600MHz} \\
\hline
\mbox{Max \# of cores/job} & 1728 & 4096\\ \hline
\mbox{Max Flops/node} & \mbox{960 GFlops} & \mbox{346 GFlops} \\ \hline
\mbox{Network} & \mbox{FDR InfiniBand} & \mbox{FDR InfiniBand} \\
\, & \mbox{Hybrid Fat-Tree} & \mbox{2-level Fat-Tree} \\ \hline
\mbox{MPI Library} & \mbox{MVAPICH 2.1} & \mbox{Intel 2015.2.164}\\ \hline
\mbox{Compiler} & \mbox{Intel 2015.2.164} & \mbox{Intel 2015.2.164}\\ \hline
\mbox{OS} & \mbox{CentOS 2.6.32-573} & \mbox{CentOS 2.6.32-431}\\ \hline
\end{array}$}
\caption{System hardware and software configuration used for Comet (\url{http://www.sdsc.edu/support/user_guides/comet.html}) and Stampede (\url{https://portal.tacc.utexas.edu/user-guides/stampede}).\label{tab:sysinfo}}
\end{table}
We note that although the systems are similar in design, Comet has newer hardware (Haswell AVX2-vectorization processors, more and faster RAM, etc) but on a standard allocation, Stampede allows runs to be performed with more than twice as many processors as Comet.  

\section{Validation}
\label{sec:valid}
\begin{figure}[htbp]
\centering
$\begin{array}{rccccc}
\rotatebox{90}{\;\;\;\;\mbox{BE+PCG (PC2)}}&
\hbox{\includegraphics[height=1.5in]{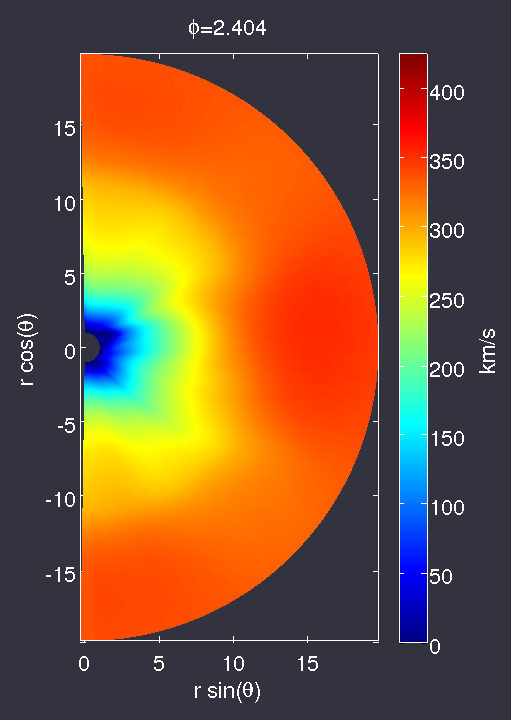}}&
\hbox{\includegraphics[height=1.5in]{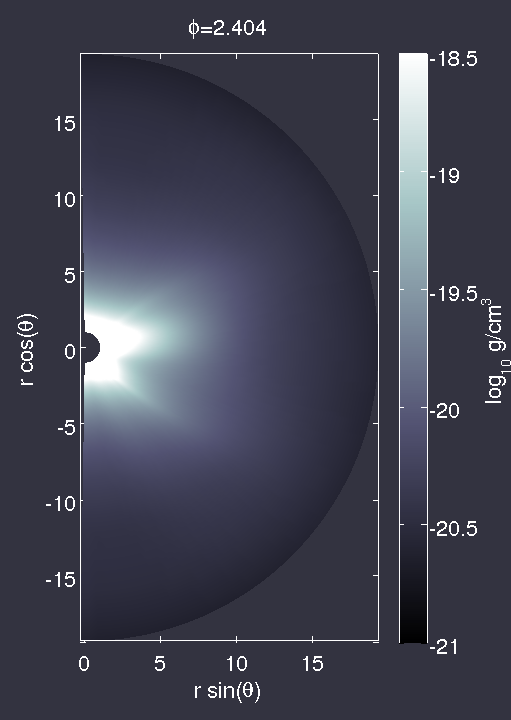}}&
\hbox{\includegraphics[height=1.5in]{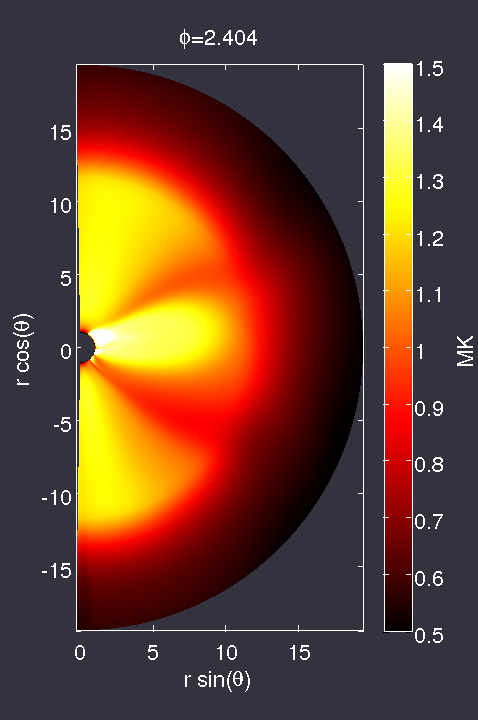}}&
\hbox{\includegraphics[height=1.5in]{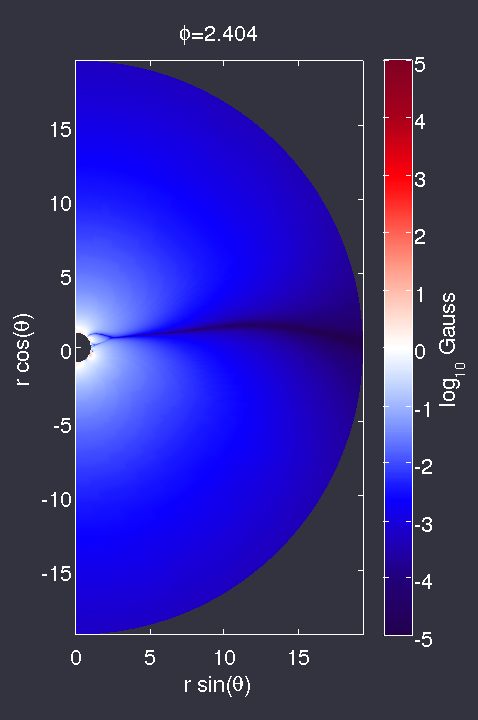}}&
\hbox{\includegraphics[height=1.5in]{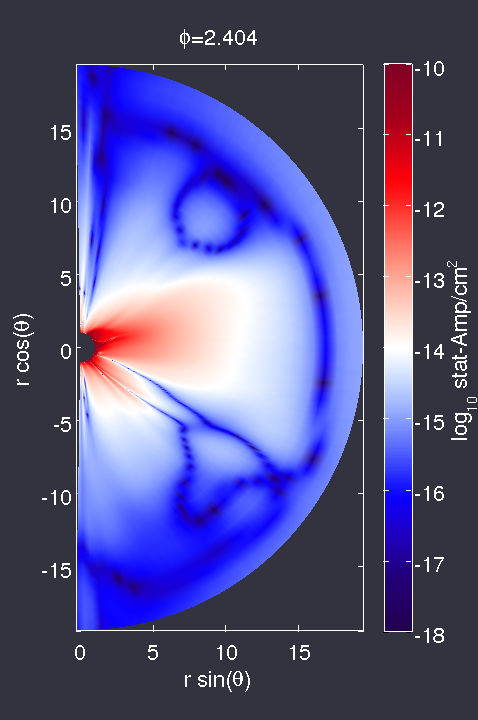}}
\\
\rotatebox{90}{\;\;\;\;\;\;\;\;\;\;\;\;\;\;\mbox{RKL2}}&
\hbox{\includegraphics[height=1.5in]{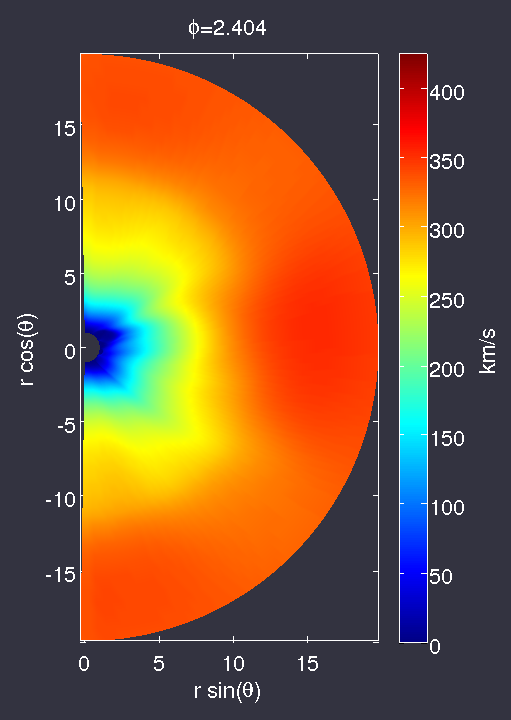}}&
\hbox{\includegraphics[height=1.5in]{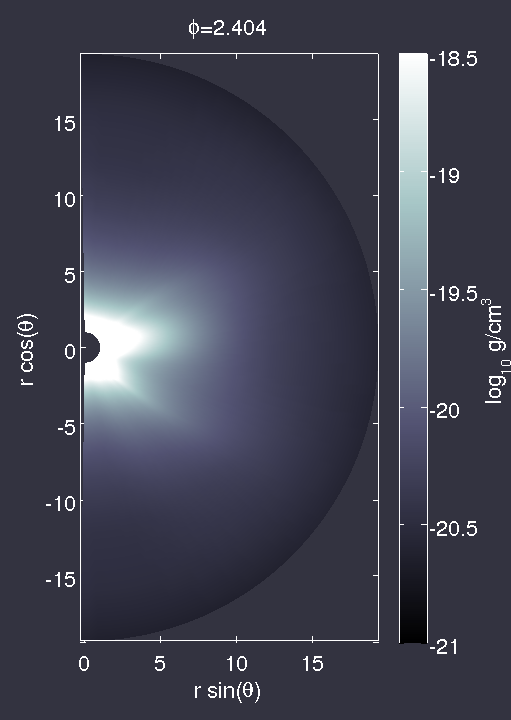}}&
\hbox{\includegraphics[height=1.5in]{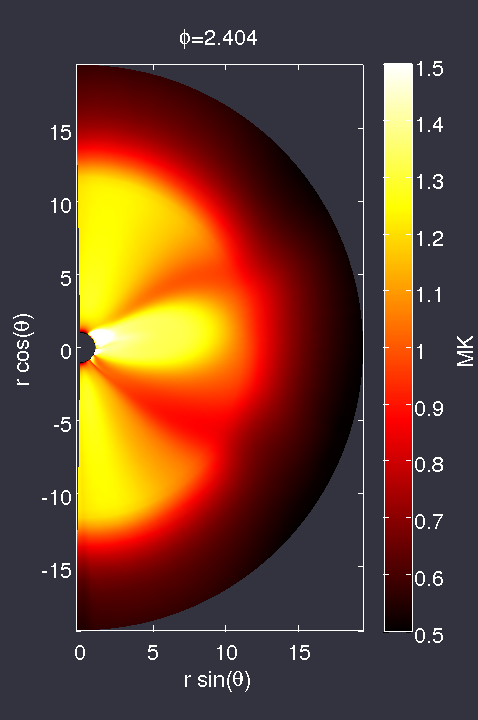}}&
\hbox{\includegraphics[height=1.5in]{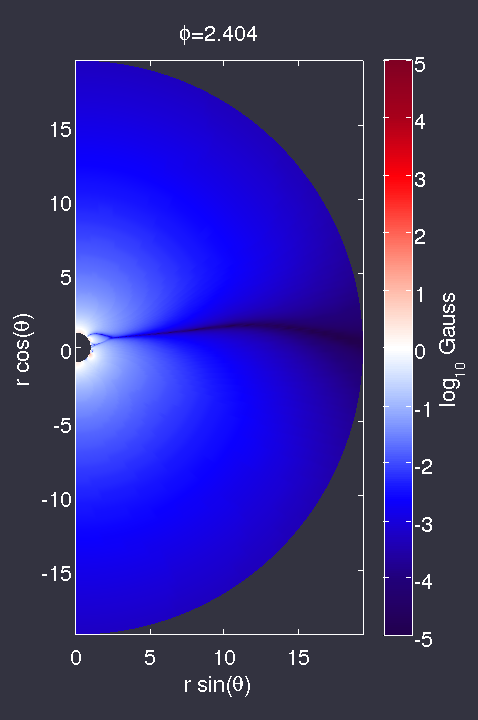}}&
\hbox{\includegraphics[height=1.5in]{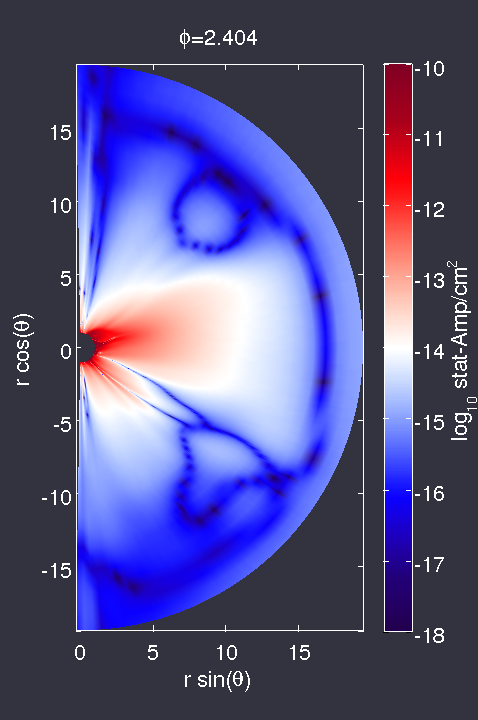}}
\end{array}$
\caption{Simulation comparions of selected fields after an 8-simulation-hour relaxation.  Left to right: $v_r$, $\mbox{log}_{10}(\rho)$, $T$, $\mbox{log}_{10}(B_r)$, and $\mbox{log}_{10}(J_{\phi})$. Top: BE+PCG (PC2), Bottom: RKL2.\label{fig:valid}}
\end{figure}

Since our focus is on comparing the two methods (BE+PCG and RKL2) to each other, we do not analyze the solution accuracy of the original BE+PCG method.  As mentioned in Sec.~\ref{sec:masmodel}, the MAS code has been used extensively and its accuracy validated \cite{CCMCcomp}.  Instead, we focus on comparing the two methods to each other by using a long simulation ($t=8$ simulation-hours) and qualitatively comparing the results.  In Fig.~\ref{fig:valid}, we show $r$-$\theta$ cuts at $\phi\approx 137^{\circ}$ of various field quantities for each run.  

\begin{figure}[htbp]
\centering
$\begin{array}{rccccc}
\; & \mbox{BE+PCG (PC2)} & \mbox{RKL2 (TC-only)} & \mbox{RKL2 (TC+Visc)} 
\\
J_{\phi}&
\raisebox{-.5\height}{\hbox{\includegraphics[width=1.5in]{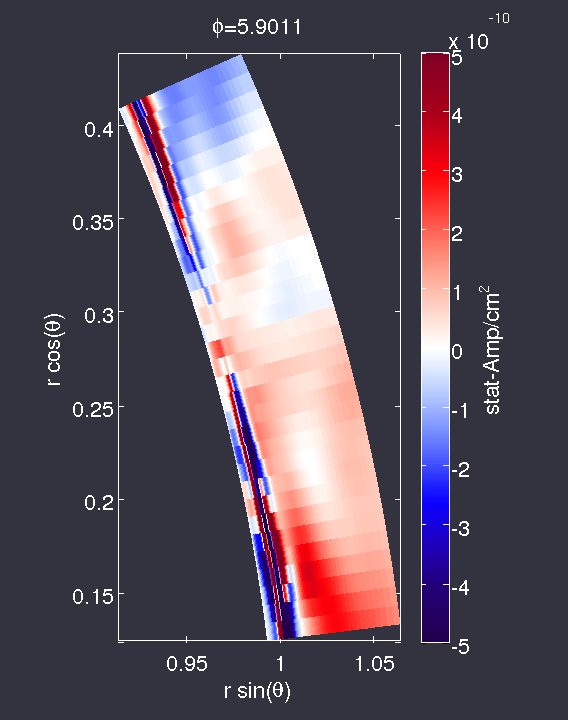}}} &
\raisebox{-.5\height}{\hbox{\includegraphics[width=1.5in]{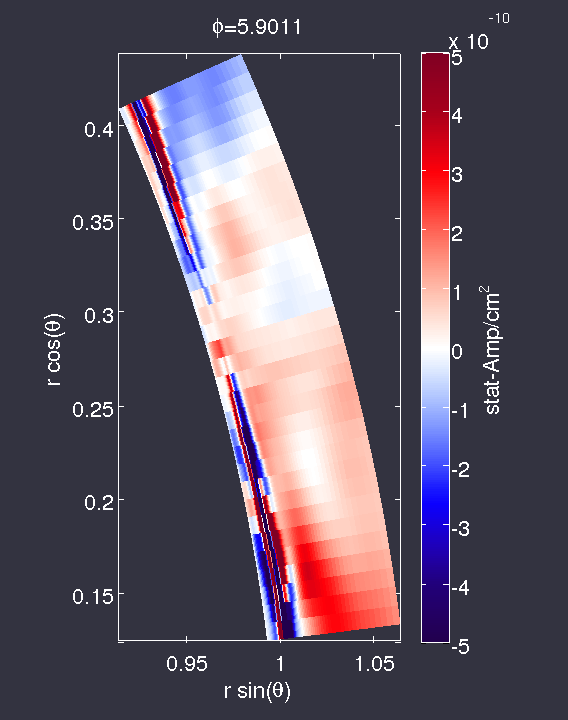}}} &
\raisebox{-.5\height}{\hbox{\includegraphics[width=1.5in]{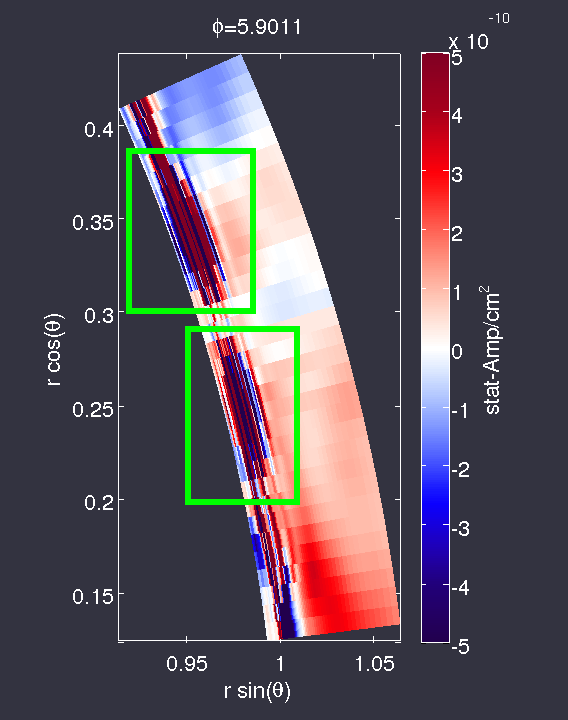}}}
\\ 
\, & \, & \, & \, & \, \\
v_{\theta}&
\raisebox{-.5\height}{\hbox{\includegraphics[width=1.5in]{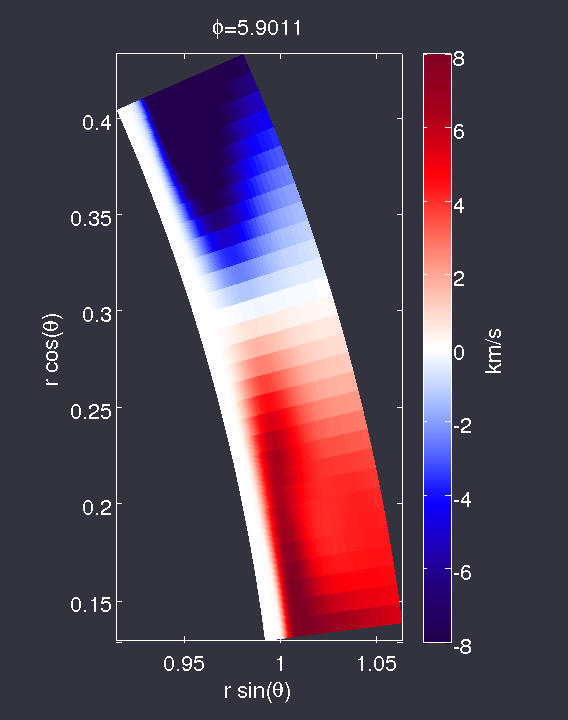}}} &
\raisebox{-.5\height}{\hbox{\includegraphics[width=1.5in]{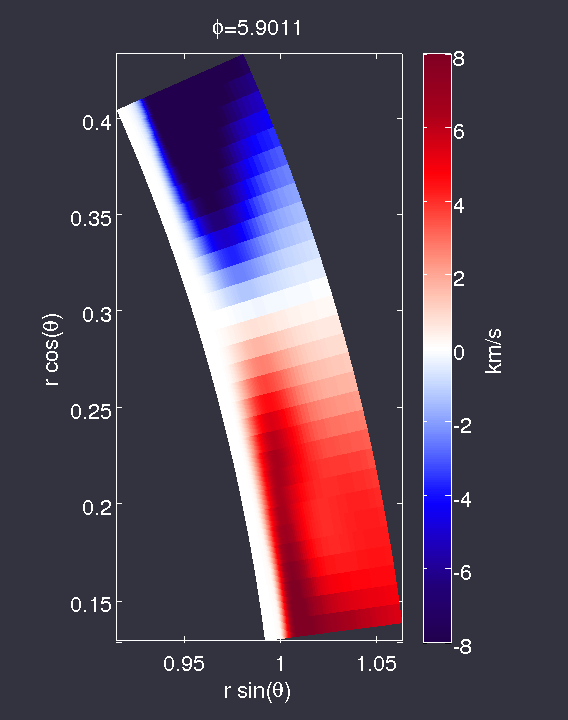}}} &
\raisebox{-.5\height}{\hbox{\includegraphics[width=1.5in]{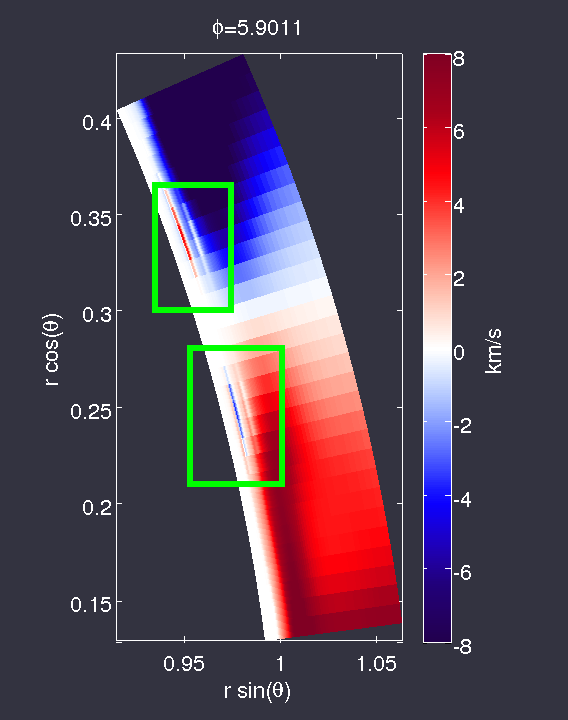}}}
\\
\, & \, & \, & \, & \, \\
T&
\raisebox{-.5\height}{\hbox{\includegraphics[width=1.5in]{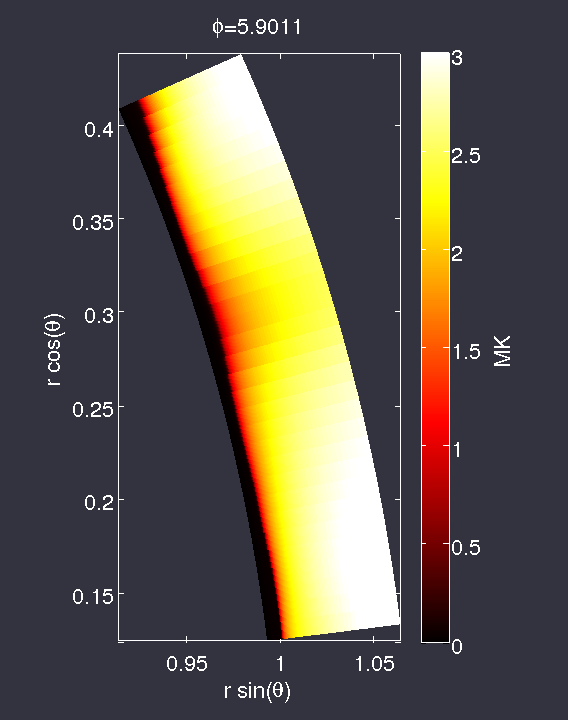}}} &
\raisebox{-.5\height}{\hbox{\includegraphics[width=1.5in]{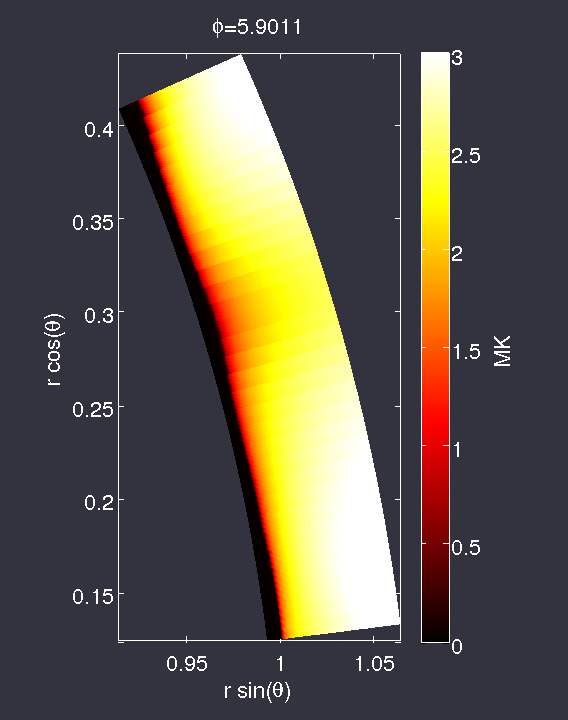}}} &
\raisebox{-.5\height}{\hbox{\includegraphics[width=1.5in]{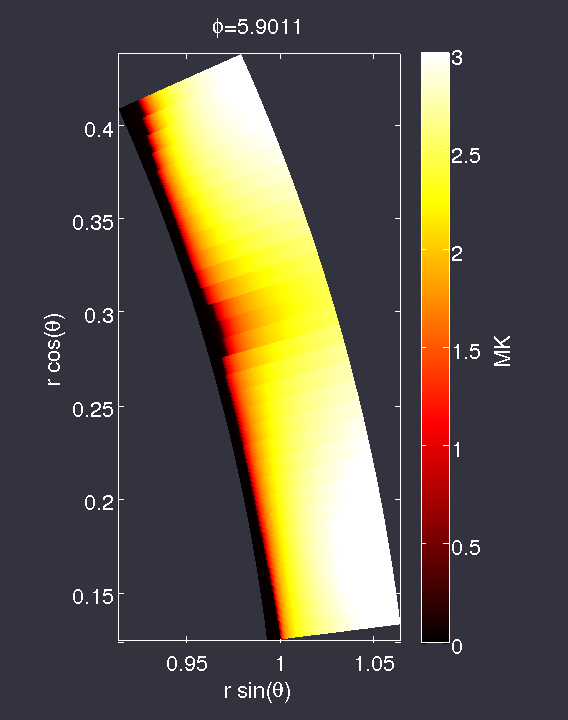}}}
\end{array}$
\caption{Simulation comparions of $J_{\phi}$, $v_{\theta}$, and $T$ near the transistion region within an active region. Left-to-right: the result using BE+PCG, RKL2 for thermal conduction only, RKL2 for both thermal conduction and viscosity.  Oscillations observed in the RKL2 case are indicated by the green boxes.\label{fig:validbad}}
\end{figure}

We see that the two methods produce very similar global results even with complicated structures present.  However, as much of the grid is focused near the transition region at roughly $r=1.02\,R_{\odot}$, it is difficult to compare the solution in those regions when viewing the full domain.   On closer inspection near the solar surface (Fig.~\ref{fig:validbad}), we once again see general agreement between the two methods, but also see an alarming difference in some localized regions where the RKL2 method exhibits high wave-mode oscillations in some fields.

A possible explanation of these oscillations is that they are exhibited due to the RKL2's inability to adequately damp high wave modes within one large time-step (as shown below).  Due to the extremely difficult conditions in transition-region physics, the standard algorithms in MAS sometimes exhibit oscillatory behavior if the spatial structures are not fully resolved.  In order to combat this effect, the artificial viscosity term is added to damp out such oscillations (more than the numerical viscosity due to the algorithm is capable), and smooth out the solution.  The true viscosity of the solar corona is so small, this term would not have been included otherwise.  In contrast to this, the thermal conduction term is an essential part of the physics of the model.  Indeed, if we run the simulations using RKL2 only for the thermal conduction (i.e. using the original BE+PCG for viscosity), the solution becomes visually identical to the BE+PCG case as shown in Fig.~\ref{fig:validbad}.

\begin{figure}[htbp]
\centering
\includegraphics[width=3in]{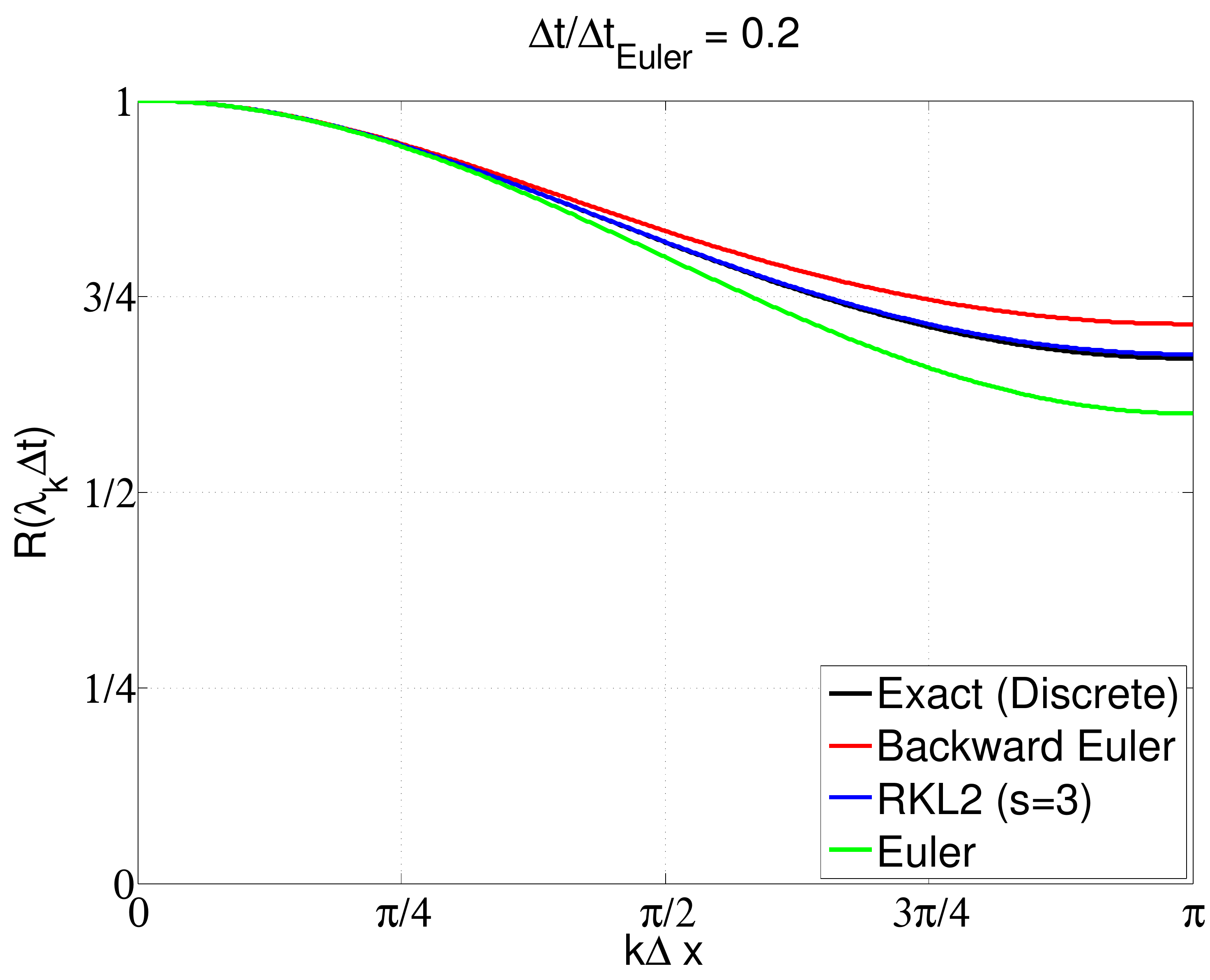}
\includegraphics[width=3in]{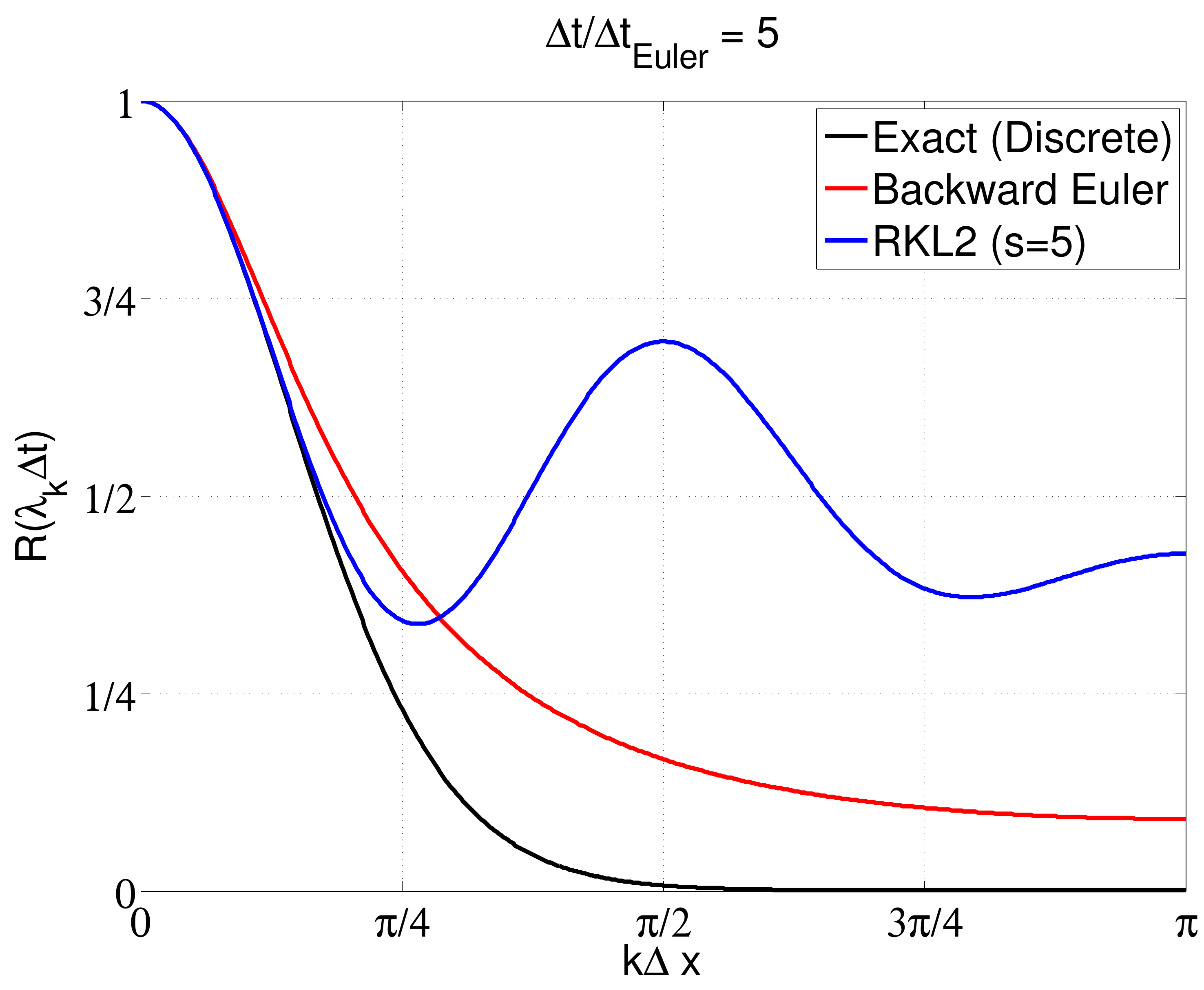}
\includegraphics[width=3in]{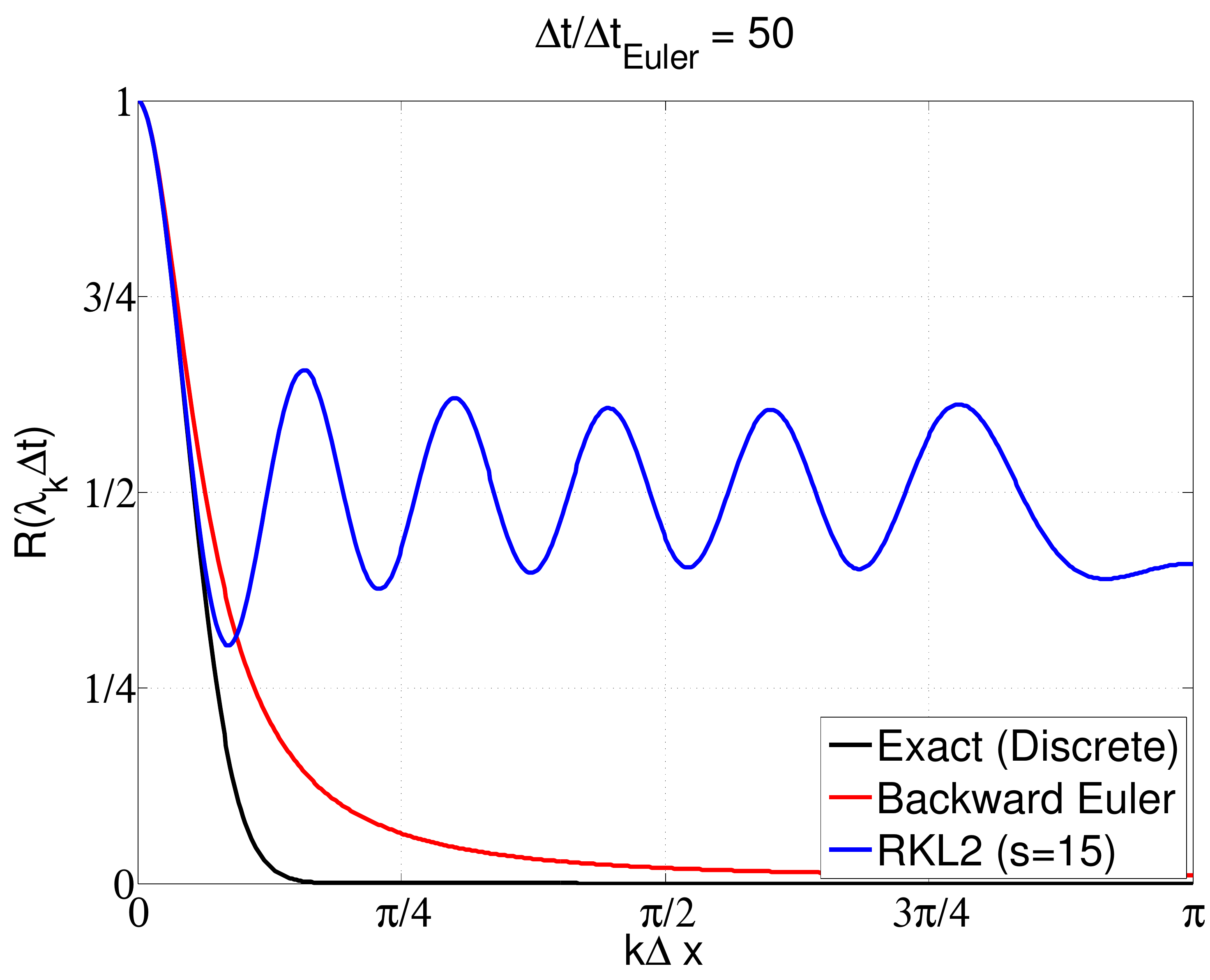}
\includegraphics[width=3in]{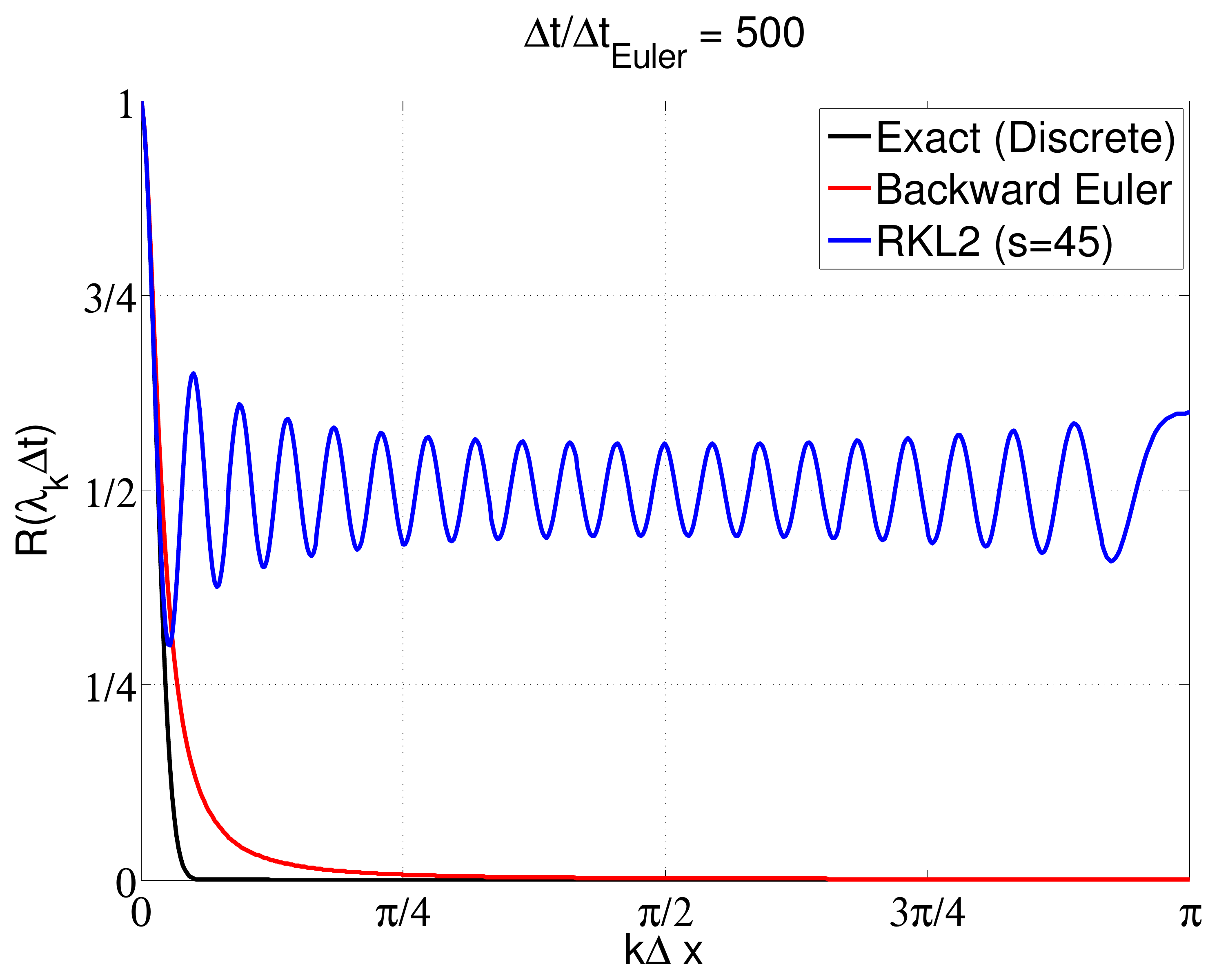}
\caption{Amplification factors for the BE and RKL2 methods for various values of $\Delta t/\Delta t_{\mbox{\tiny Euler}}$.  The exact discrete solution is shown for comparison, and the explicit Euler method is shown for the case where it is stable.\label{fig:ampfac}}
\end{figure}

To gain some insight into why the RKL2 method does not damp out the high wave modes, we first note that even though the amplification factor of a convergent scheme is less than 1 (i.e. it is stable and  diffusive), it is possible that it is too large to properly damp out a growing wave mode.  In Fig.~\ref{fig:ampfac}, we show the amplification factors for the BE and RKL2 methods for the one-dimensional case of Eq.~(\ref{eq:pde}) with a constant diffusion coefficient discretized on a uniform grid with spacing $\Delta x$.  The factors are derived from the analytic eigenvalues of the system and are shown for integrating at a time-step $1/5$, $5$, $50$, and $500$ times the explicit Euler limit $\Delta t_{\mbox{\tiny Euler}}$ (in our simulations, viscosity is being integrated at over $700\times$ the limit in some portions of the grid, and thermal conduction over $10000\times$).  The amplification factor for the exact solution of the semi-discrete system and for the explicit Euler method (in the case where it is stable) are shown for comparison. 

We first notice that for small wave modes, the RKL2 is much more accurate than BE, which is not surprising as it is a second-order accurate method.  We also see that when exceeding the Euler limit by a large factor, the amplification factors for both BE and RKL2 are much larger than the true solution, and the RKL2's amplification factor is much larger than the BE's.  Also, the RKL2's amplification factor does not tend towards 0 as $k\Delta x\rightarrow \pi$ implying that the RKL2 method is generally not L-stable, although it appears to be strongly A-stable \cite{PDEBOOK} and is positive preserving.  The large amplification factor of the RKL2 for high modes will not damp high-frequency solutions efficiently within large time-step because modes which should be virtually eliminated according to Eq.~(\ref{eq:pde}) will only be damped by a factor of $\sim0.5$.  However, with successive application of the method, the wave modes will eventually damp.  Therefore, if Eq.~(\ref{eq:pde}) were integrated by itself, the RKL2 would be convergent (given enough steps and total simulation time), but when coupled to a larger multi-physics model, the modes may not be damped.    We note that BE is not exempt from this issue as its amplification factor is many times larger than the true solution, but since its amplification factor tends towards 0 for high modes, and is generally small, it is expected to be more robust than RKL2.  

To support this conclusion, we have tested sub-cycling the RKL2 method within each large time-step, and find that it yields a solution visually identical to the BE method given enough sub-cycles ($\sim 10$ in our case).  We also found that sub-cycling explicit Euler steps for $\sim 1/4$ of the time-step and completing the remainder of the step with RKL2 also yielded correct results.  This is understandable as the explicit Euler method damps the high modes very efficiently such that running it for $\sim 1/4$ of a time-step damps high modes more than even a full step with BE.  These considerations may lead to new scheme strategies that combine methods in order to get both an efficient and overall correct solution.

This analysis indicates that the RKL2 (and other STS methods) may be inadequate for use in multi-physics codes when the parabolic operator is being used either for smoothing oscillations and/or shocks (like artificial viscosity) and perhaps even if they are being used in a balanced equation where the diffusion must counter high-mode growth due to reaction or other driving terms.  

The amplification factors shown in Fig.~\ref{fig:ampfac} for RKL2 are qualitatively different than those shown in Ref.~\cite{RKL2_2012} and Ref.~\cite{RKL2_2014}.  This is because the factors shown there are for the RKL methods where the number of sub-steps $s$ is chosen first, and the time-step factor is then set exactly from the equation for $s$ (such as Eq.~(\ref{eq:rkl2_s})).  In such an exact case, the RKL2's amplification factor \emph{does} tend towards zero for very high modes, but still remains near $\sim 1/2$ for a large portion of the frequency spectra.  In a real-world case where the time-step is set by other factors, and then the minimum stable $s$ is computed (the suggested procedure given in Ref.~\cite{RKL2_2014}), the amplification factors become like those shown in Fig.~\ref{fig:ampfac}.  Even one extra $s$ step over the exact amount needed causes the qualitative change.  Interestingly, using any extra number of sub-steps for the \emph{RKL1} scheme eliminates the problem shown in Ref.~\cite{RKL2_2014} of it approaching magnitude 1 at $k\Delta x=\pi$, possibly making that scheme more useful than otherwise assumed (although it is not positive preserving).

For uniform grid methods, if one could compute $\Delta t_{\mbox{\tiny Euler}}$ exactly, one could modify the time-step of the system to exactly follow Eq.~(\ref{eq:rkl2_s}) for the nearest whole value of $s$ and potentially mitigate this problem.  However, if computing $\Delta t_{\mbox{\tiny Euler}}$ exactly is not efficiently possible (which is often the case) and/or one is using non-uniform grids, the number of sub-steps $s$ will always be more than exact, at least somewhere on the grid.  

Even though our solutions exhibit the oscillations shown in Fig.~\ref{fig:validbad}, since they are limited to a few localized regions and do not seem to effect the rest of the global solution (as shown in Fig.~\ref{fig:valid}), we still report the timing results for the runs as a comparison of the scaling of the methods.  Since our main concern in the timings is parallel scaling, we note that any modifications along the lines described above to help the RKL2 method damp the high modes (e.g. sub-cycling Euler steps) may increase the total compute time, but should not effect the scaling, as those steps have the same communication profile as the RKL2 method itself.  Also, since using the RKL2 for thermal conduction alone produces a consistent solution, the timing results for thermal conduction have increased relevance.

\section{Timing Results}
\label{sec:timings}
To time the simulations, we use calls to the MPI library's timing API routines.  It is well known that massive parallel codes can have a very wide range of run times when using many processors due to fluctuations of system load and communication occupancy.  In order to extract the scaling out of the methods in as an ideal situation as possible, we use multiple runs (between 4 and 6) of each processor count and take the best result.  In addition to recording the total time for each portion of the run tested (thermal conduction and viscosity), we also record the `non-compute' time for each portion which we define as the time the algorithms are involved in inter-processor communication (including buffer loading and unloading, data transfer, and synchronization/waiting).  

\begin{figure}[htbp]
\centering
\scalebox{0.9}{$\begin{array}{|l|l|l|l|}
\hline
\multicolumn{2}{|c|}{{\bf \mbox{Comet}}} & \multicolumn{2}{|c|}{{\bf \mbox{Stampede}}}\\ \hline
N_{\mbox{\tiny total}} & (N_r,N_{\theta},N_{\phi}) & N_{\mbox{\tiny total}} & (N_r,N_{\theta},N_{\phi})\\ \hline
24 & (2,3,4) & 64 & (2,4,8) \\ \hline
48 & (2,4,6) & 128 & (4,4,8) \\ \hline
96 & (3,4,8) & 256 & (4,8,8) \\ \hline
216 & (4,6,9) & 512 & (4,8,16) \\ \hline
432 & (6,8,9) & 1024 & (8,8,16) \\ \hline
864 & (6,8,18) & 2048 & (8,16,16) \\ \hline
1796 & (6,12,24) & 4096 & (8,16,32) \\ \hline
\end{array}$}
\raisebox{-.5\height}{
\includegraphics[width=0.5\textwidth]{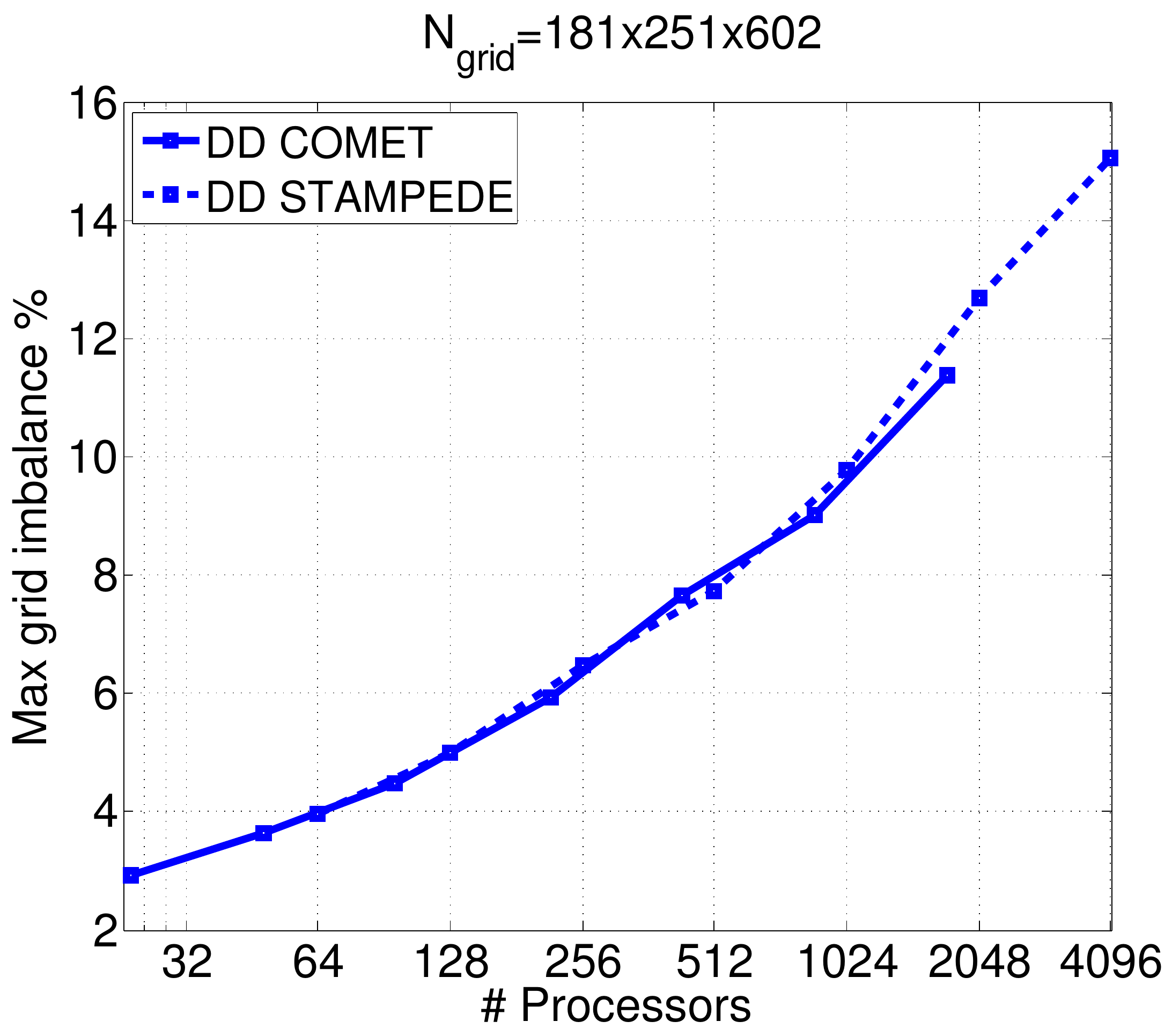}}
\caption{Left: Total cores and processor topology use for domain decomposition (DD) on Comet and Stampede.  Right: Maximum possible grid-based load imbalance due to DD.\label{fig:procs}} 
\end{figure}

Since Comet and Stampede each have different number of processor cores per node, the domain decomposition is unique for runs on each system, along with the number of total processors for each run.  On Comet, the runs are performed using 1 node up to the maximum of 72 nodes, while on Stampede, due to the smaller amount of system memory per node, the simulations could not be run on fewer than 4 nodes. The domain is split up along processors in each direction yielding 3D block sub-domains.  Since the grid size in each dimension will typically not divide evenly into the processor count along that dimension, some cores will have one extra grid point along some dimensions than others.  In the worst case, cores may have one extra grid point along all three dimensions.  For a fixed problem size, this causes an intrinsic load imbalance that increases as the number of total processors increases.  In Fig.~\ref{fig:procs} we show the processor topologies used on each system and the maximum grid-based load imbalance.  
An additional possible source of load imbalance comes from cores containing sub-domains along physical boundaries, which have more or less work to perform based on the boundary conditions.  For example, in the case of the viscosity solver, the cores containing sub-domains along the polar boundaries require calculating a polar average involving communication/synchronization with all other cores conditioning the polar boundaries.  

\begin{figure}[p]
\centering
$\begin{array}{rcc}
\; & \mbox{Thermal Conduction} & \mbox{Viscosity}
\\
\rotatebox{90}{\mbox{Comet}} &
\raisebox{-.5\height}{\hbox{\includegraphics[width=0.375\textwidth]{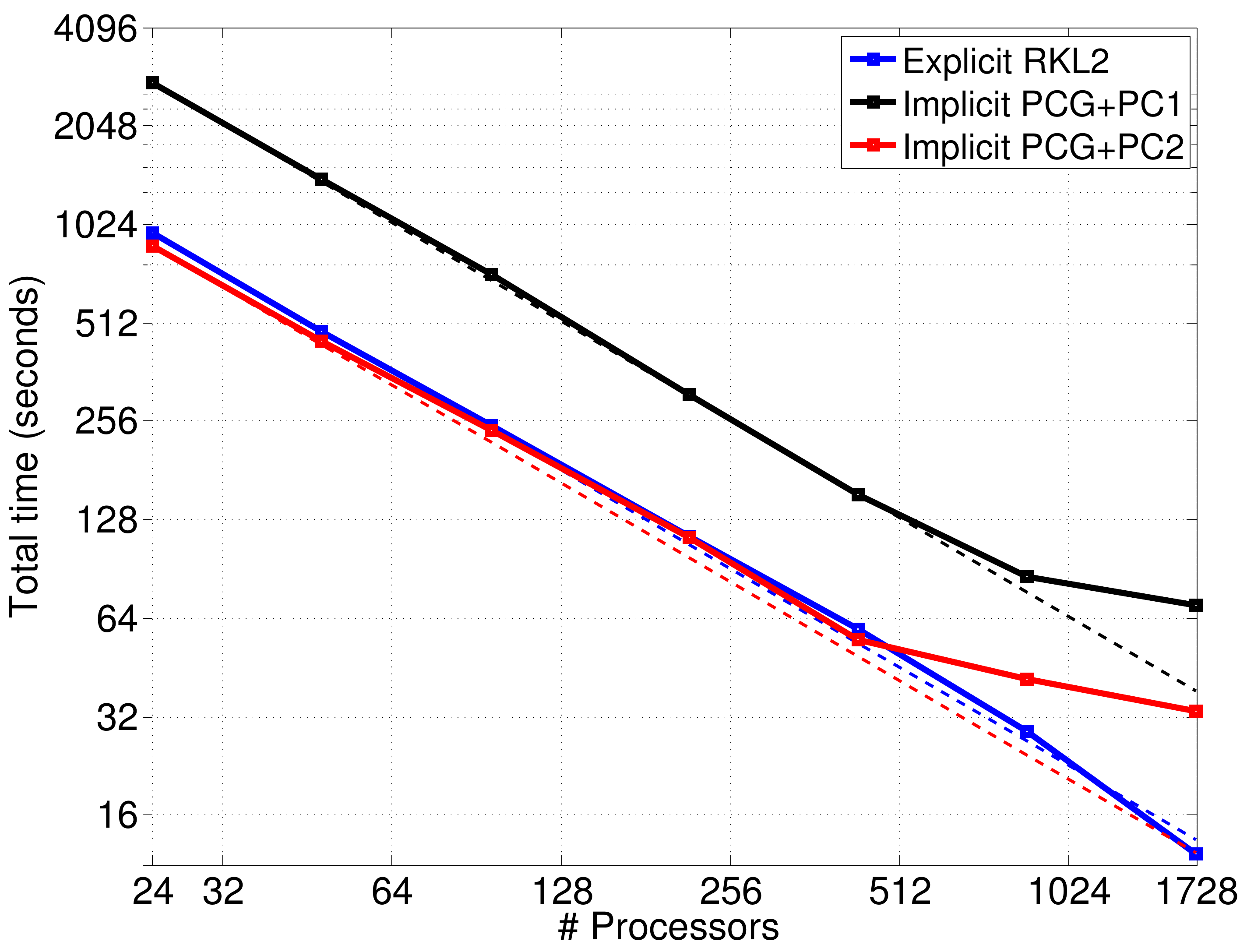}}} & 
\raisebox{-.5\height}{\hbox{\includegraphics[width=0.375\textwidth]{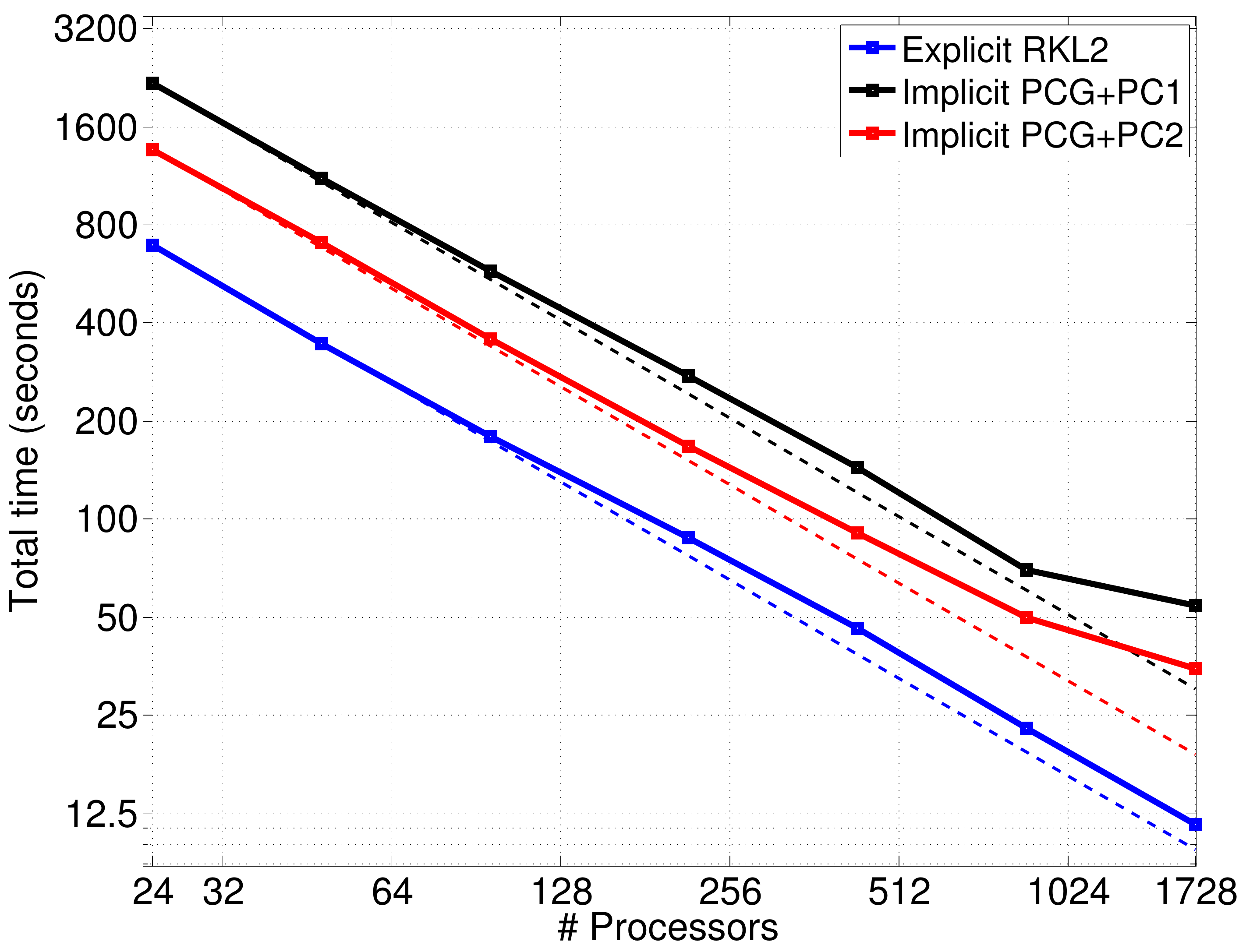}}}
\\
\rotatebox{90}{\mbox{Stampede}} &
\raisebox{-.5\height}{\hbox{\includegraphics[width=0.375\textwidth]{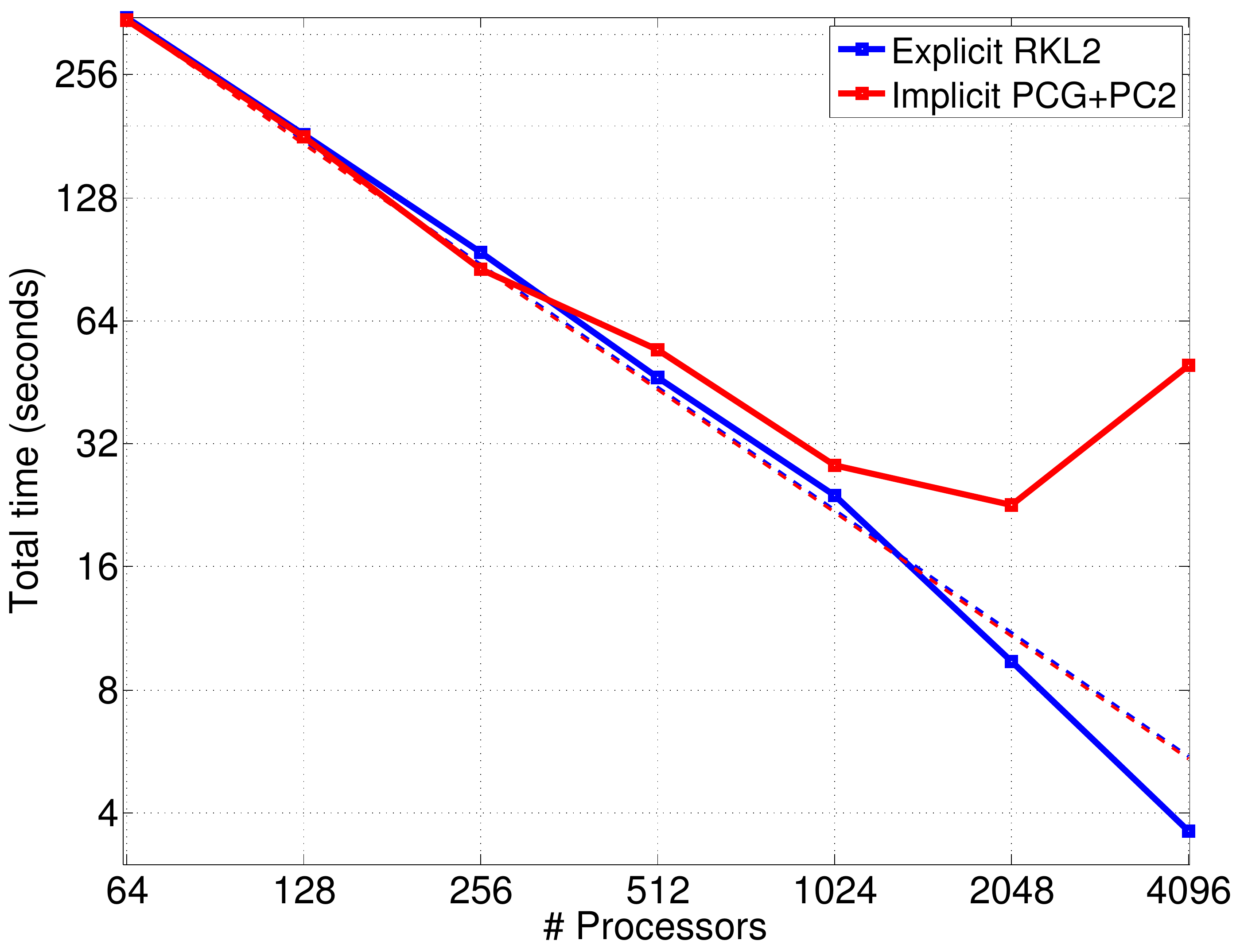}}} &
\raisebox{-.5\height}{\hbox{\includegraphics[width=0.375\textwidth]{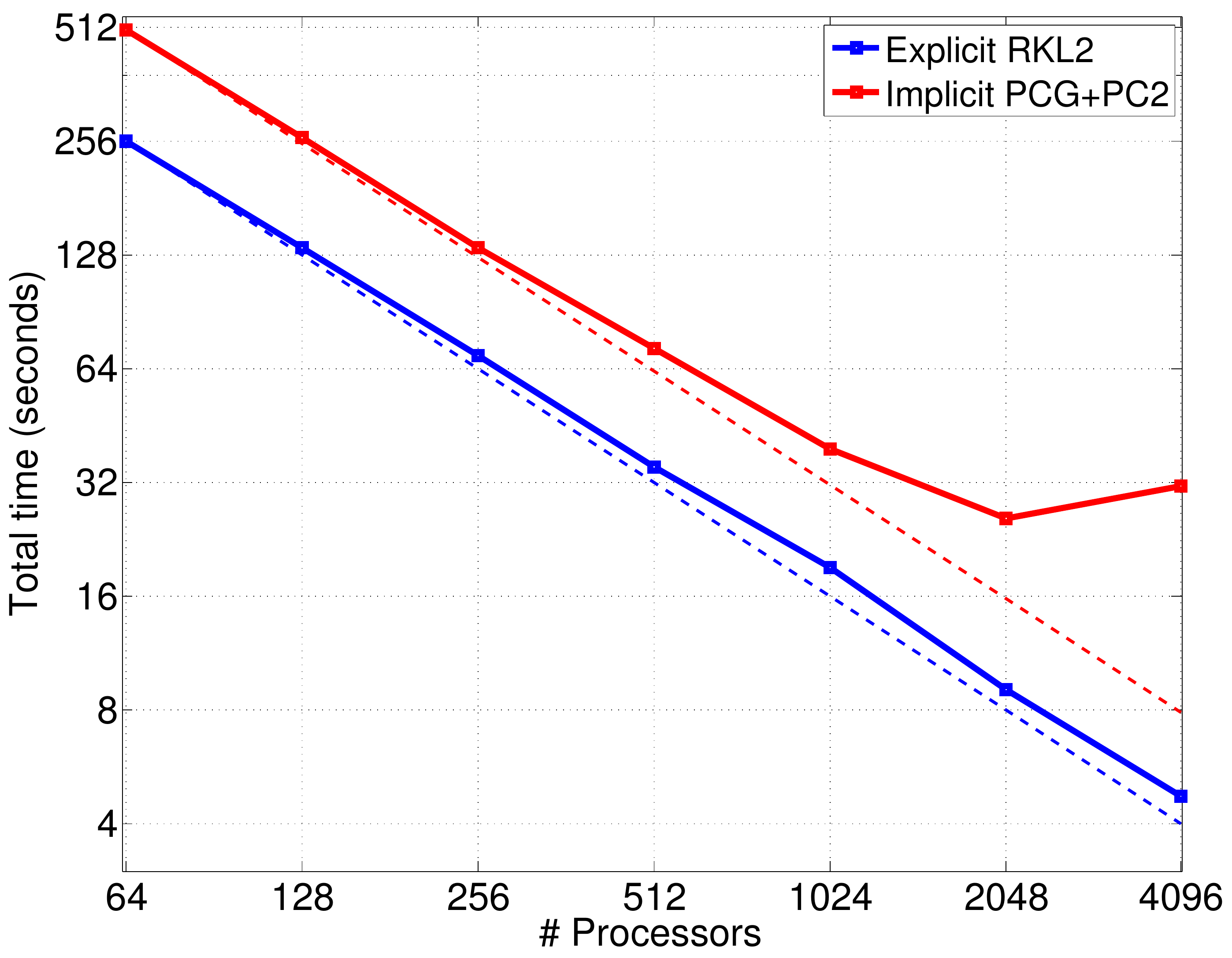}}}
\end{array}$
\caption{Timing results for the thermal conduction (left) and viscosity (right) portions of the simulations computed on Comet (top) and Sampede (bottom). \label{fig:timings}} 
\end{figure}

The timing results for thermal conduction and viscosity are shown in Fig.~\ref{fig:timings}.  Since the BE+PCG simulations with PC1 were found to be much slower than either BE+PCG PC2 or RKL2 during the Comet runs, they were not repeated on Stampede.  

\begin{wrapfigure}{r}{0.5\textwidth}
\centering
\includegraphics[width=0.4\textwidth]{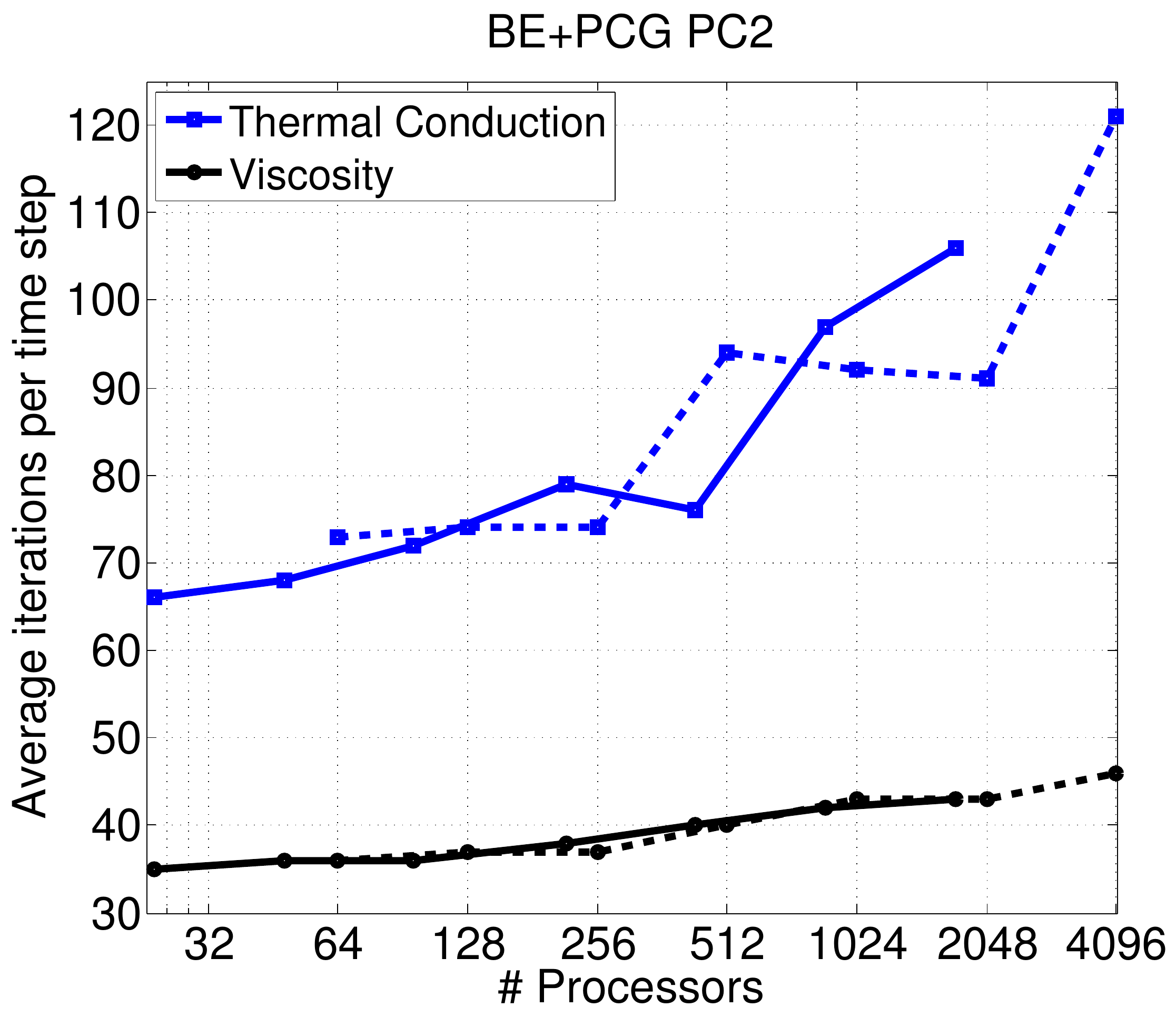}
\caption{Average number of iterations for the BE+PCG PC2 method applied to thermal conduction and viscosity for runs on Comet (solid line) and Stampede (dashed line).  The RKL2 method takes an average of 195 sub-steps for thermal conduction and 49 for viscosity. \label{fig:iters}} 
\end{wrapfigure}

We first observe that all methods scale well up to 512 processors.  Higher then that, we see that the BE+PCG method starts to loose scaling, and in some cases, the run time starts to increase with larger number of cores.  The BE+PCG PC1 seems to scale a bit better than PC2, but is overall a much slower method.  Therefore, if the RKL2 method is within the desired accuracy, in this case it is shown to perform far better than BE+PCG PC1 (which is still a commonly used method).  The RKL2 is shown to scale better than the BE+PCG PC2 method in all cases.  For thermal conduction, the run time for RKL2 is very similar to the BE+PCG PC2 up to the point that the scaling starts to diminish.  This is so even though it is using many more sub-steps than the number of iterations of BE+PCG PC2 (see Fig.~\ref{fig:iters}).  This shows that one sub-step of the RKL2 method is much faster than one iteration of the PCG PC2.  

Using the maximum number of processors on a standard allocation, the thermal conduction using RKL2 exhibited a speedup of over $2.5$X for Comet and $13.7$X for Stampede compared to the BE+PCG PC2 method.  For viscosity, the RKL2 is at least twice as fast as BE+PCG PC2 for all processor counts, and due to its better scaling, reaches a maximum speed increase over BE+PCG PC2 of $\approx 3$X on Comet and over $6.5$X on Stampede.

Another interesting observation is that it appears that the thermal conduction operator does not scale as well as the viscosity operator when using BE+PCG.  One possible reason for this is that the number of iterations required for the PCG solver increases as the processor count increases (due to the degradation of the PC2 {--} see Sec.~\ref{sec:PC}) much more so for thermal conduction than it does for viscosity.

\begin{figure}[p]
\centering
$\begin{array}{rcc}
\; & \mbox{Thermal Conduction} & \mbox{Viscosity}
\\
\rotatebox{90}{\mbox{Comet}} &
\raisebox{-.5\height}{\hbox{\includegraphics[width=0.375\textwidth]{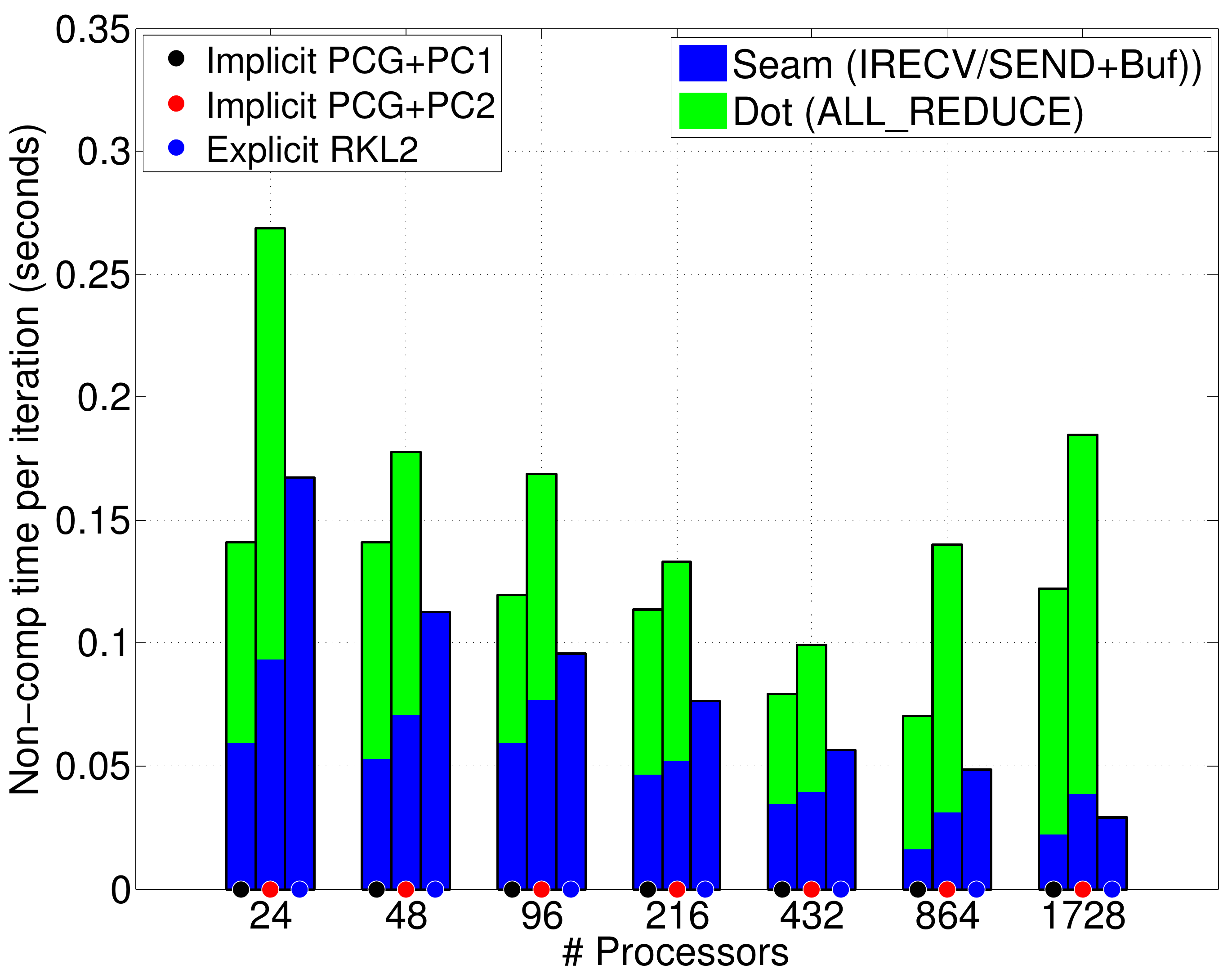}}} & 
\raisebox{-.5\height}{\hbox{\includegraphics[width=0.375\textwidth]{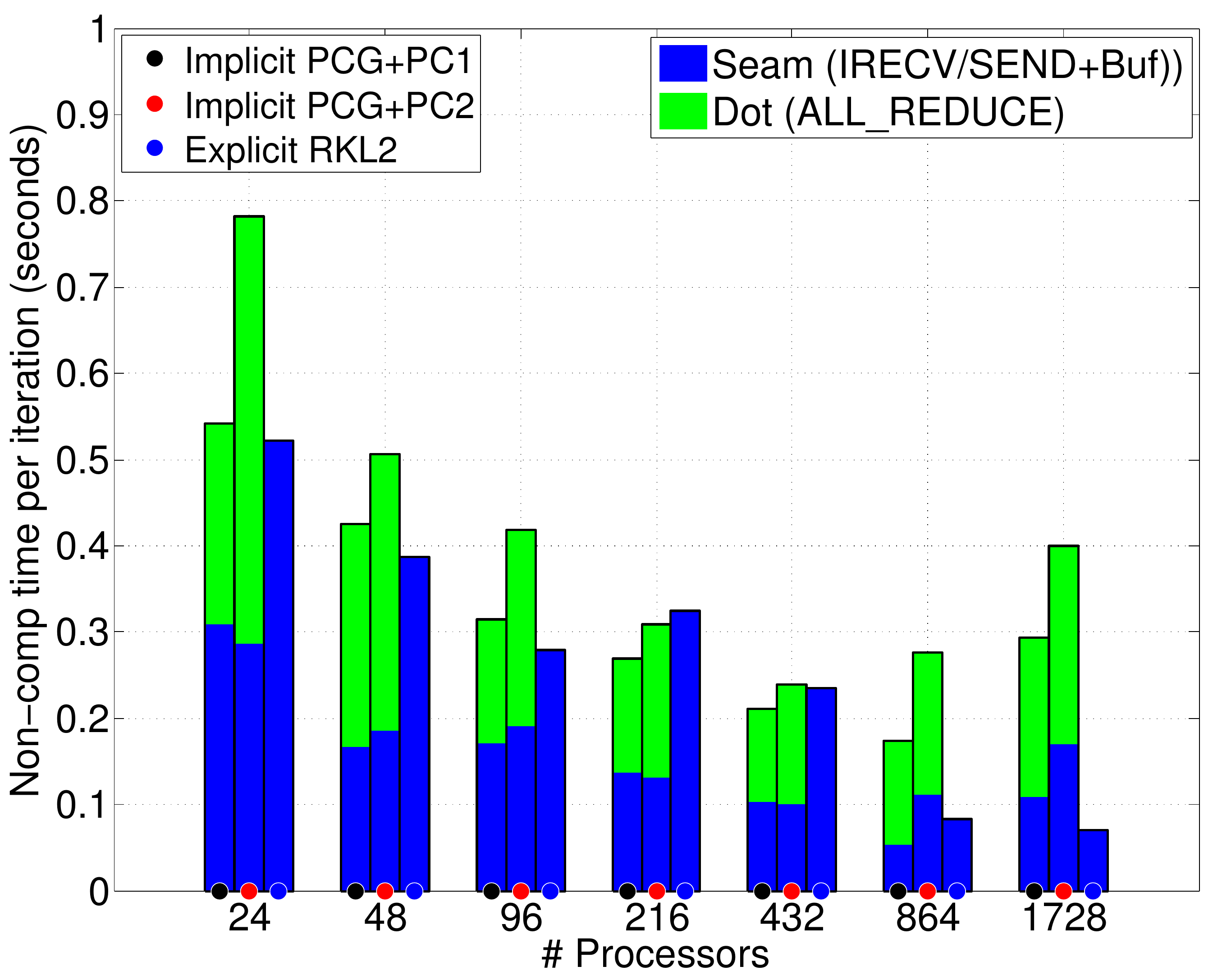}}}
\\
\rotatebox{90}{\mbox{Stampede}} &
\raisebox{-.5\height}{\hbox{\includegraphics[width=0.375\textwidth]{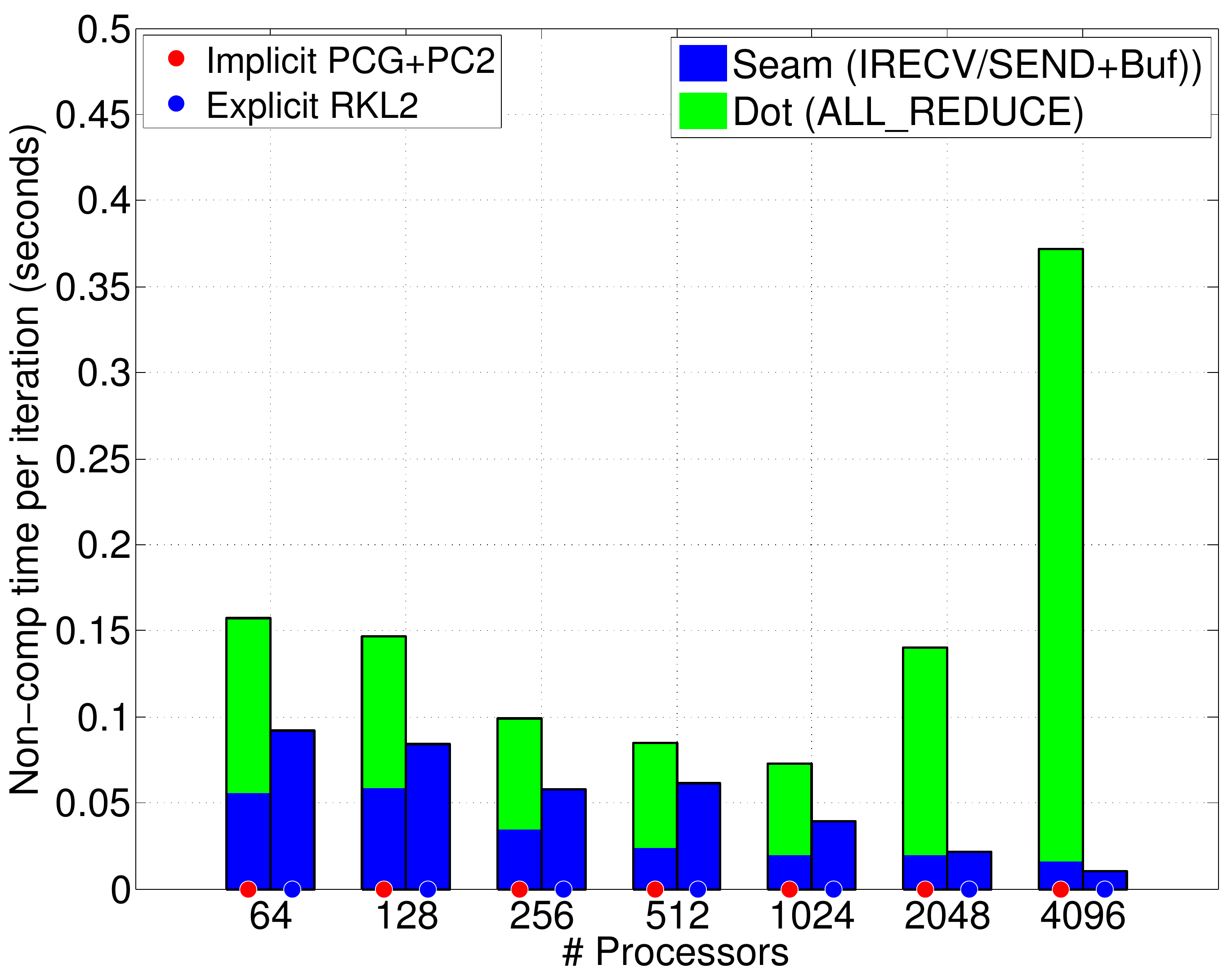}}} &
\raisebox{-.5\height}{\hbox{\includegraphics[width=0.375\textwidth]{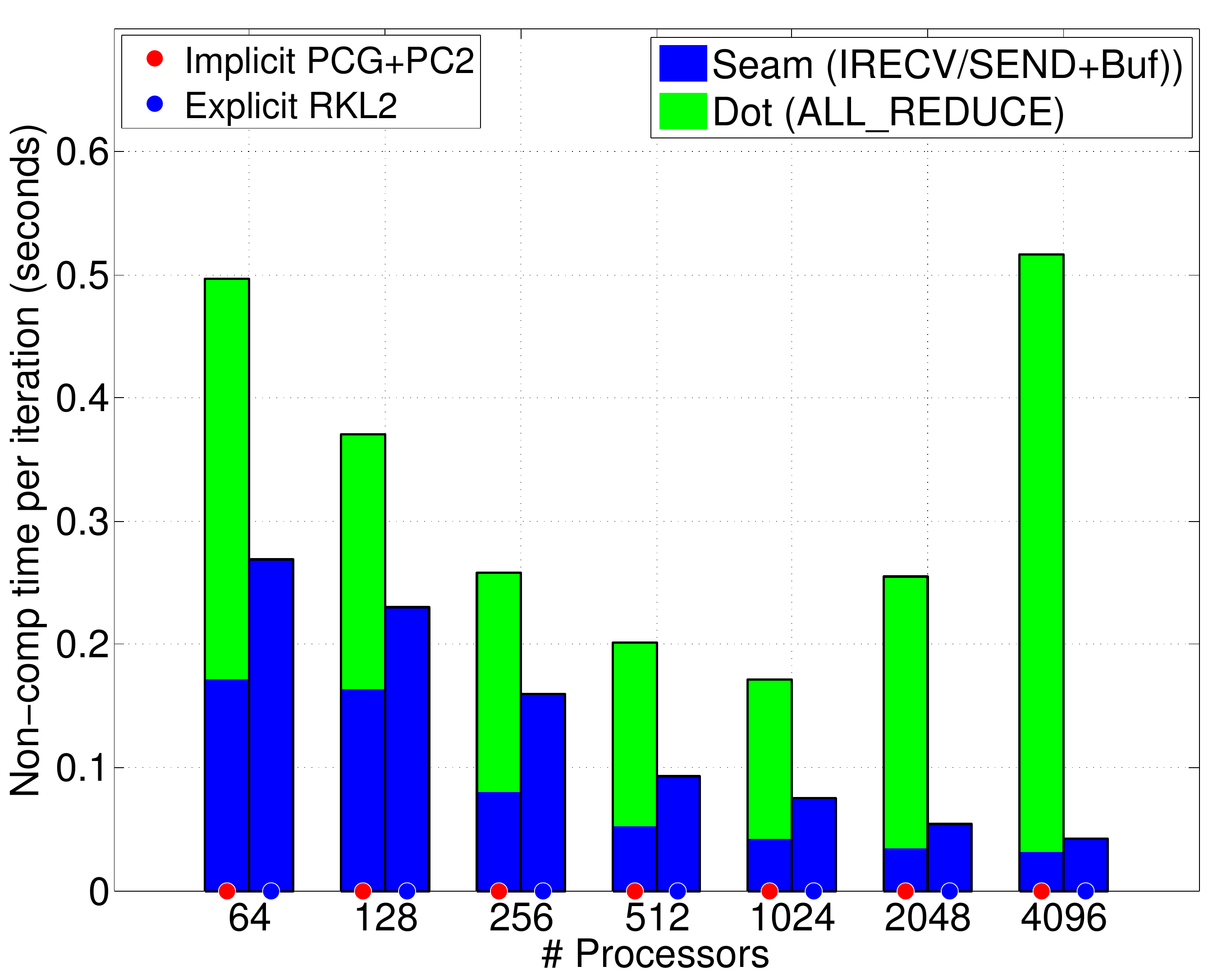}}}
\end{array}$
\caption{Non-compute time per iteration/sub-step for thermal conduction (left) and viscosity (right) portions of the simulations computed on Comet (top) and Stampede (bottom).  The method used for each bar is indicated by the colored dots along the horizontal axis.  \label{fig:noncomp}} 
\end{figure}

It is informative to explore the communication (or `non-compute') time of the runs, which we display (per iteration/sub-step) in Fig.~\ref{fig:noncomp}.  We have separately recorded the non-compute time for both local (peer-to-peer) and global communications.  We see that the peer-to-peer non-compute time for the RKL2 method is somewhat higher than for the PCG methods (which is unexpected as they both use the same matrix-multiply routines). As the processor count increases, this non-compute time decreases (due to the reduction in the number of transferred data points) as expected.  However, for the PCG methods, the global non-compute time (which is always larger than the point-to-point time) stays relatively constant for a while, but then starts to increase as the processor count gets high.  For Stampede, this effect is the greatest.  Since at the time these runs were performed, Stampede was much more oversubscribed then comet, the network congestion could be a cause of this drastic increase.  In either case, we see that the RKL2's lack of global communications is key to its improved scaling.  Through testing using MPI barriers, we have found that it is not the AllReduce MPI routine itself in the dot products that is the main cause of the PCG's lack of scaling, but rather it is the routine's global synchronization acting as a barrier, causing all imbalances encountered before the routine to `pile up'.  As the amount of computational work per processor decreases, the effect of these imbalances increases.   

\begin{figure}[htbp]
\centering
$\begin{array}{|l|c|}
\hline 
\multicolumn{2}{|c|}{\mbox{Time Breakdown from Relaxation}}\\
\hline
\mbox{Semi-implicit Pred+Corr} & 46\% \\
\mbox{\color{blue} Viscosity} & {\color{blue} 16\%} \\
\mbox{\color{blue} Thermal Conduction} & {\color{blue} 14\%} \\
\mbox{Alfv\'en Waves} & 19\% \\
\mbox{Resistivity} & 1\% \\
\mbox{Rest of code} & 4\% \\ \hline
\end{array}$
\raisebox{-.5\height}{
\includegraphics[width=0.5\textwidth]{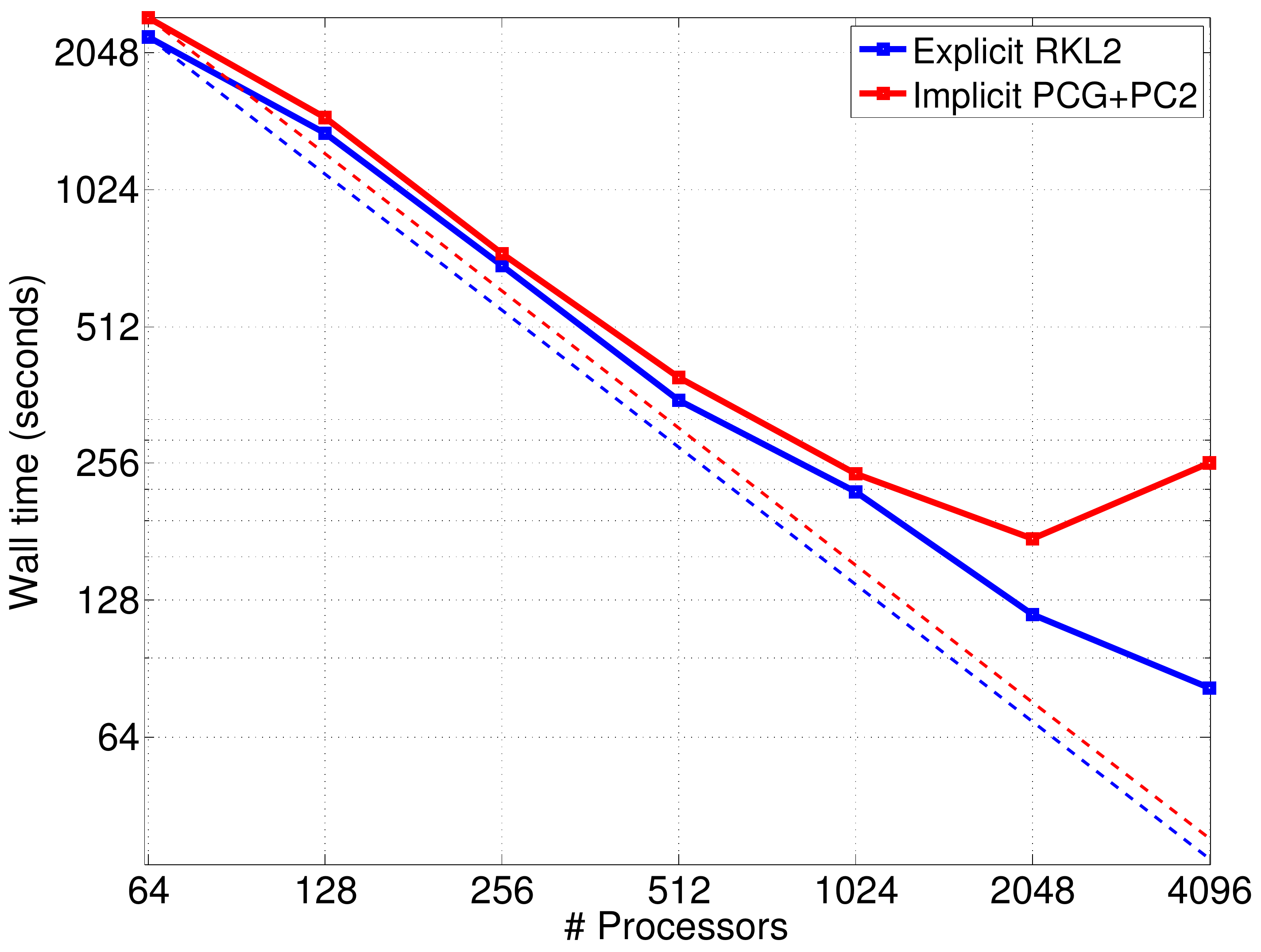}}
\caption{Left: Breakdown of the computational cost of each section of the MAS code for the validation run of Sec.~\ref{sec:valid} using BE+PCG PC2.  Right: Total wall time for the timing runs computed on Stampede for the two methods.  Even though the two parabolic operators tested only accounted for $30\%$ of the wall time in the BE+PCG PC2 runs, the wall time when using RKL2 on 4096 cores is reduced by over 3X.\label{fig:wall}} 
\end{figure}

Finally, although our focus is on the timing of the thermal conduction and viscosity operators, in Fig.~\ref{fig:wall} we show the total wall time of the MAS code for the Stampede runs of Fig.~\ref{fig:timings} and what percentage of the total run time each part of the MAS code contributed to (based on the validation runs of Sec.~\ref{sec:valid}).
It is interesting to note that although the parabolic operators used to test the methods only accounted for $\sim 30\%$ of the code's run time when using BE-PCG PC2, the effect of using RKL2 seems to improve scaling to the extent that the overall code exhibits a speedup of over 3X for 4096 cores on Stampede.  Why this is the case is still unclear, but may have to do with easing the load on the infiniband buffers, and/or allowing the load imbalance to equalize/catch-up between operators.

\section{Conclusions}
\label{sec:conclusion}
In this paper we have compared a representative algorithm from two classes of methods for integrating parabolic operators past the standard explicit time-step limits in the context of thermodynamic MHD simulations of the solar corona.  The first method is an implicit scheme solved with a Krylov solver: the backward-Euler scheme with the preconditioned conjugate gradient (BE+PCG) with either a diagonal (PC1) or incomplete LU (PC2) preconditioner.  The other is a super time-stepping method: the second-order Runge-Kutta Legendre (RKL2) scheme.   The methods were described in detail and tested on a production-level coronal relaxation simulation of the described MAS code.  Special attention was given to the performance and scaling of the methods. 

We found for our specific case, that the RKL2 scales much better for high core counts than the BE+PCG methods, and matches or exceeds the performance for low numbers of cores.  For integrating viscosity, the RKL2 method performed over twice as fast as the BE+PCG.  In other tests (not reported here), the speedup of the RKL2 method was even larger (over 3X).  The global synchronization of the dot products in the BE+PCG method were found to be the most likely cause of the poor scaling at high core counts, since any load imbalance would be bottle-necked there.

Although the RKL2 method has superior performance and scaling, we found that, when used for viscosity, it exhibited slowly-growing grid-level oscillations in a few localized portions of the domain.  An analysis of the methods showed that these oscillations (generated by other parts of the model due to the presence of unresolved structures) are not adequately damped by the RKL2 method due to its high amplification factor for large wave modes.  Some methods of avoiding these errors were discussed, but each would slow down the algorithm.  Using RKL2 for only the thermal conduction operator yielded correct results, implying that its use may be limited to specific use-cases.  Our analysis also showed that the BE+PCG method, while not immune to the same problem, is much more robust due to it being  L-stable.  

Future work will include searching for ways to modify the RKL methods to alleviate the damping issues discussed above.  As for the Krylov solvers, investigation into new non-blocking PCG methods has the potential to improve their scaling.  Another interesting direction would be to formulate hybrid schemes, which combine the advantages of the multiple methodologies.  Research into the application of STS-type methods to wave equations is of particular interest, as it could yield a scalable alternative to semi-implicit operators.

\section*{Acknowledgments}
Work at Predictive Science was supported by AFOSR, NASA, and NSF.  Computational resources at the Texas Advanced Computing Center (TACC) and the San Diego Supercomputer Center (SDSC) were provided by the NSF-supported XSEDE program.  The ADAPT map was produced collaboratively between AFRL and NSO/NISP.  

\section*{References}
\bibliographystyle{iopart-num}
\bibliography{Caplan_ASTRONUM16_STS}

\section*{Appendix A}
The following table depicts the pseudo-code for formulating and applying the diagonal (PC1) and ILU0 (PC2) preconditioners used for testing the BE+PCG method of Sec.~\ref{sec:BEPCG}.
\begin{center}
\begin{tabular}{|l|c|c|} \hline
\, & Formulation (${\bf P}$)        & Application ($z={\bf P}^{-1}\,r$)
\\ \hline
\;     & \;                  &\;              
\\
PC1     & $\begin{array}{l}{\tt do \;i=1:N} \\\qquad P_{ii}=A_{ii} \\{\tt enddo}\end{array}$
&
$\begin{array}{l}{\tt do \;i=1:N} \\ \qquad z_i=r_i/P_{ii}\\ {\tt enddo}\end{array}$     
\\ \hline
PC2     & $\begin{array}{l}
\qquad \notag \\
{\bf LU}={\bf A} \\
{\tt do\; i=2:N}\notag  \\
\qquad {\tt do\; k=1:i-1}\qquad (LU_{ik}\ne0)  \\
\qquad \qquad  LU_{ik}=LU_{ik}/LU_{kk}\notag  \\
\qquad \qquad {\tt do\; j=k+1:N} \qquad (LU_{ij}\ne0)   \\
\qquad \qquad \qquad LU_{ij}=LU_{ij}-LU_{ik}\,LU_{kj}   \\
\qquad \qquad {\tt enddo} \\
\qquad {\tt enddo} \\
{\tt enddo \qquad \qquad}\notag \\
{\bf P}={\bf LU} \\
\qquad
\end{array}$   &  $\begin{array}{l}  
{\tt do\,i=1:N} \\
\qquad z^*_i=r_i \\
\qquad {\tt do\,j=1:i} \qquad (LU_{ij}\ne 0) \\
\qquad \qquad z^*_i=z^*_i-LU_{ij}\,z^*_j \\
\qquad {\tt enddo} \\
{\tt enddo} \\
{\tt do\, i=N:1} \\
\qquad z_i=z^*_i \\
\qquad {\tt do\, j=i+1:N} \qquad (LU_{ij}\ne 0)  \\
\qquad \qquad z_i=z_i-LU_{ij}\,z_j \\
\qquad {\tt enddo} \\
\qquad z_i=z_i/LU_{ii}\\
{\tt enddo}
\\
\,
\end{array}
$ 
\\ \hline
\end{tabular}
\end{center}
\newpage
\section*{Appendix B}
\label{sec:dtexp}
In order to compute the minimum required number of sub-steps for the RKL2 method of Sec.~\ref{sec:rkl2}, one requires a (preferably close) estimate for the Euler stability time-step:
\begin{equation}
\label{eq:dteuler}
\Delta t_{\mbox{\tiny Euler}}=2/|\lambda|_{\mbox{\tiny max}},
\end{equation}
where $|\lambda|_{\mbox{\tiny max}}$ is the largest absolute value eigenvalue of the matrix ${\bf M}$ of Eq.~(\ref{eq:BE_lin}).  For finite difference methods, an upper bound on $|\lambda_{\mbox{\tiny max}}|$ can be given as 
\[
|\lambda_{\mbox{\tiny max}}| < \mbox{max}(\alpha\,{\tilde k}^2),
\]
where $\alpha$ is the diffusion coefficient, and $\tilde k$ is the inverse discrete grid element (e.g. for spherical coordinates, ${\tilde k}^2\approx 4\,[1/\Delta r^2 + 1/(r\,\Delta \theta)^2 + 1/(r\,\sin(\theta)\,\Delta \phi)^2]$.  However, this often leads to a very large under-estimate of  $\Delta t_{\mbox{\tiny Euler}}$, especially when using anisotropic diffusion.

A much better estimate can easily be made by the sometimes overlooked method of using Gershgorin disk theorem \cite{GERSHBOOK} as was done in Ref.~\cite{RK4STB}.  A Gershgorin disk is defined as
\begin{deff}
Given a square matrix ${\bf A}$, a Gershgorin disk for every row $j$ 
is defined as a disk in the complex plane, centered at $A_{jj}$, with a radius of: 
\[
R_j=\sum_{i\neq j}^N\,|A_{i,j}|.
\]
\end{deff}
and the following theorem applies to ${\bf A}$:
\begin{thm}
Every eigenvalue $\lambda$ of a square matrix ${\bf A}$ lies in one of its Gershgorin disks:
\[\left|\lambda-A_{jj}\right|\leq R_j\].
\end{thm}
Since in our case, $\lambda\in \mathcal{R}$ and $A_{jj}\in\mathcal{R}$, we get a bound on $|\lambda_j|$ for each row $j$ as $|{A}_{jj}|-R_j\leq |\lambda_j| \leq |{A}_{jj}|+R_j$.   Therefore the maximum value of $|\lambda|$ is bound by:
\begin{equation}
\label{eq:gstb}
|\lambda|_{\mbox{\small max}} \leq \mbox{max}\left\{\sum_{i=1}^N\,|A_{i,j}|,\forall j \,\mbox{rows},\right\}.
\end{equation}
Since the matrix coefficients are already explicitly formed for use in the solvers/integrators, Eq.~(\ref{eq:gstb}) is  trivial to calculate.

The stability bound on $\Delta t_{\mbox{\tiny Euler}}$ found by using Eq.~(\ref{eq:gstb}) is often much closer to the true bound than the method using ${\tilde k}$ above.  The accuracy of the bound for real symmetric matrices was shown in Ref.~\cite{Gbound} to be within a factor of $\sqrt{p}$ of the true bound, where $p$ is the maximum number of non-zero values in the rows of ${\bf A}$.  For the cases tested in this paper, $p=15$ for viscosity and $p=19$ for thermal conduction.  However, we note that for many cases of finite difference operators, the Gershgorin bounds can be much closer to the true bounds than the maximum error of $\sqrt{p}$ implies \cite{RK4STB}.
\end{document}